\documentclass[12pt]{article}

\usepackage{titling}
\newcommand{\subtitle}[1]{%
  \posttitle{%
    \par\end{center}
    \begin{center}\large#1\end{center}
    \vskip0.5em}%
}

\usepackage{times}
\usepackage{xcolor}
\usepackage{booktabs}
\usepackage{longtable}
\usepackage{multirow}
\usepackage{subcaption}
\usepackage{graphicx}
\usepackage{amsfonts}
\usepackage{amsmath}
\usepackage{soul}
\usepackage{hyperref}
\usepackage{setspace}
\usepackage[capitalise]{cleveref}  
\usepackage[labelfont=bf]{caption}
\usepackage{xurl}

\usepackage[final]{changes}

\newcommand{\ignore}[1]{}
\newcommand{\abr}[1]{\textsc{#1}}

\newcommand{\newsguard}{NewsGuard}

\newenvironment{sciabstract}{%
\begin{quote} \bf}
{\end{quote}}

\newcounter{lastnote}

\title{Using co-sharing to identify use of mainstream news for promoting potentially misleading narratives}

\author
{Pranav Goel,$^{1\ast}$ Jon Green,$^{2}$ David Lazer,$^{3,4}$ Philip S. Resnik$^{5,6}$\\
\\
\normalsize{$^{1}$Network Science Institute, Northeastern University, Boston, MA, USA}\\
\normalsize{$^{2}$Department of Political Science, Duke University, Durham, NC, USA}\\
\normalsize{$^{3}$Network Science Institute, Northeastern University, Boston, MA, USA}\\
\normalsize{$^{3}$Institute for Quantitative Social Science, Harvard University, Cambridge, MA, USA}\\
\normalsize{$^{5}$Department of Linguistics, University of Maryland, College Park, MD, USA}\\
\normalsize{$^{6}$Institute for Advanced Computer Studies, University of Maryland, College Park, MD, USA}\\
\\
\normalsize{$^\ast$To whom correspondence should be addressed; E-mail: p.goel@northeastern.edu}
}

\date{}

\begin{document} 


\baselineskip24pt


\maketitle 



\begin{sciabstract}

Much of the research quantifying volume and spread of online misinformation measures the construct at the source level, identifying a set of specific unreliable domains that account for a relatively small share of news consumption. 
This source-level dichotomy obscures the potential for users to repurpose factually true information from reliable sources to advance misleading narratives. 
We demonstrate this potentially far more prevalent form of misinformation by identifying articles from reliable sources that are frequently co-shared with (shared by users who also shared) ``fake" news on social media, and concurrently extracting narratives present in fake news content and claims fact-checked as false. 
Specifically in this study, we use Twitter/X data from May 2018 to November 2021 matched to a U.S. voter file. 
We find that narratives present in misinformation content are significantly more likely to occur in co-shared articles than in articles from the same reliable sources that are not co-shared, consistent with users using information from mainstream sources to enhance the credibility and reach of potentially misleading claims. 
\end{sciabstract}



\section*{Introduction}

There is much concern in the social sciences \replaced{about}{with} the prevalence of ``fake news'' --- defined in Lazer \emph{et al} \cite{lazer2018science} as ``fabricated information that mimics news media content in form but not in organizational process or intent'' --- in the online information environment \cite{che2018fake,weeks2021s}. 
This \textit{process-driven} conceptualization of fake news has fostered \textit{source-level} measures of misinformation more broadly. 
Sources --- and all of their constituent stories --- are categorized as reliable or unreliable based in large part on the respective sources' adherence (or lack thereof) to the procedures and norms of mainstream journalism, in addition to their reputations for veracity. 
These categorizations are then projected onto all of the information that the sources produce. 
This process is streamlined by the publication of domain lists, curated by researchers (e.g. \cite{grinberg2019fake}) or third-party vendors (e.g. \cite{newsguard}), that can be applied to any set of URLs for ease of classification \cite{allcott2017social,budak2019happened, robertson_users_2023}. 

Within this framework, information is evaluated not on the basis of its truth value, but instead on the reputation of its source. 
This provides a limited foundation for identifying and addressing the challenges misinformation poses. 
Practically speaking, ``fake" news sources capture a \replaced{small}{trivial} share of overall \replaced{news consumption}{web traffic} \cite{allcott2017social,grinberg2019fake,guess2019less,allen2020evaluating}. 
For example, one study estimated that only $1\%$ of individuals on Twitter accounted for $80\%$ of all exposure to fake news sources \cite{grinberg2019fake}, while another study concluded that over $90\%$ of individuals on Facebook did not share any fake news sources \cite{guess2019less}. 
If this constituted the sum total of false or misleading content of empirical interest, then the widespread concern regarding misinformation \cite{aral2019protecting,lovari2020spreading,ognyanova2020misinformation,kubin2021role,green2022online} would seem to be misplaced. 

\added{One practical way in which domain lists might underestimate the extent of the misinformation challenge is these lists becoming quickly outdated as the information landscape evolves continuously \cite{giglietto2023workflow}. }
However, the broader limitations of source-level measures of misinformation are theoretical. 
The binary categorization of sources as either ``fake" or ``reliable" obscures the reality that information produced by unreliable sources can be true, while information produced by reliable sources can be false or misleading. 
Moreover, the binary categorization of information as either true or false does not account for the extent to which information --- especially on the internet --- does not exist independently from its social function \cite{thorson_curated_2015}. 
Groups of users do not share individual pieces of information for information's sake alone, but instead share information to support broader claims that advance their interests. 
These claims may be less true than the sum of their parts, as users may pick and choose true pieces of information from reliable sources to advance potentially misleading narratives. 
The extent to which information informs --- rather than misinforms --- often depends on how it is used.

We demonstrate this dynamic by examining the networked nature of information on social media. 
Specifically, we examine patterns in \textit{co-sharing} \added{on Twitter} by first identifying users who share information from unreliable sources, and then examining the information from reliable sources that those users also share (while Twitter was rebranded as X in 2023, in this article we continue to refer to the platform as Twitter since our dataset predates the rebranding).
This allows us to score articles from mainstream sources based on their propensity to be co-shared with ``fake'' news. 
We then use automated tools to extract narrative structures from mainstream articles, ``fake'' news articles, and fact-checked statements known to be false. 
We expect that mainstream stories with high scores on this measure will be significantly more likely than those with low scores to contain narratives present in misinformation content, or potentially misleading narratives --- entity-action relationships that repeatedly occur in information produced by unreliable sources or occur in statements known to be false. 
This suggests a dynamic in which users seeking to promote potentially misleading narratives may use factually true information to do so, to the extent to which mainstream sources produce information that is useful for this purpose. 
This dynamic reflects recent work showing that many partisan users selectively curate politically consistent information from politically diverse sources \cite{greencuration, gonzalez2023asymmetric}, challenging source-level measures of political slant. 
It further suggests that false and misleading claims might be far more prevalent on social media than source-level measures alone suggest, highlighting the importance of adopting a measurement approach that incorporates user behavior and the interaction between information from different sources \cite{watts2021measuring}.

\begin{figure*}[!t]
\centering
\captionsetup{font=footnotesize}
\includegraphics[width=0.9\textwidth]{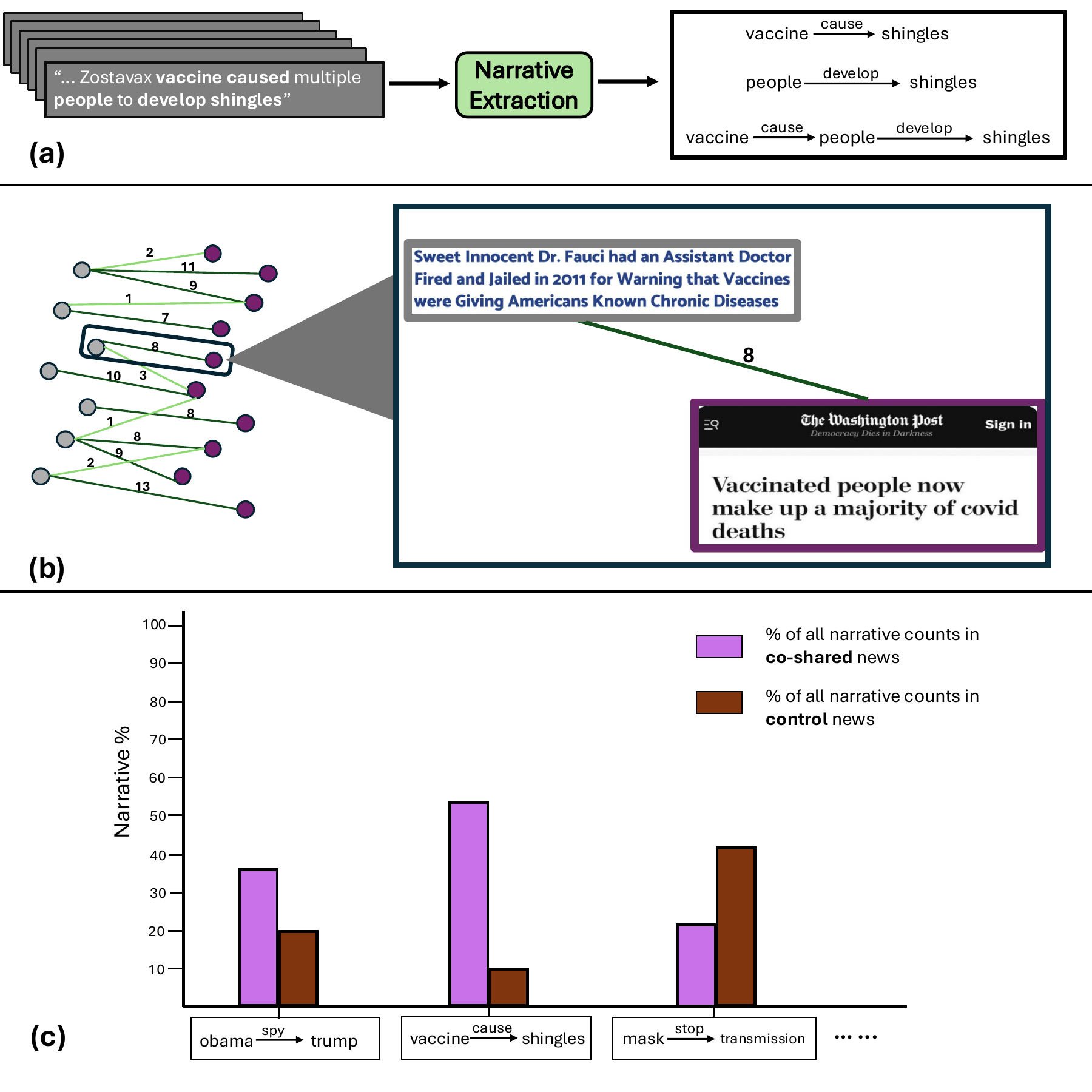}
\caption{\textbf{Visual illustration summarizing our approach.} \textbf{(a)} We automatically extract narrative structures from all texts in our data: misinformation content and mainstream news articles. Libraries of potentially misleading narratives are created using the narratives extracted from misinformation content. Narrative extraction approach is summarized in Fig. \ref{fig:narr_extraction}. \textbf{(b)} To find reliable outlet articles disproportionately co-shared with fake news, we construct a graph with edges between fake and reliable articles (the nodes) weighted by the number of users sharing both incident articles. For example, the two articles highlighted above, one by a fake news outlet and one by The Washington Post, are both shared by $8$ unique individuals on Twitter. We then score all pairs of fake and reliable news on the likelihood of being shared by the same set of Twitter users, when controlling for the individual popularity of the two articles (edges in lighter shade are below the threshold for being considered as co-shared). \textbf{(c)} For each narrative from the potentially misleading narrative library, we compute their presence in co-shared news and non-co-shared (control) news. Presence is calculated as the percentage of articles in a set (co-shared or control) that a given narrative occurs in. Via an aggregated comparison across potentially misleading narratives, we find that these narratives present in misinformation content are significantly more likely to occur in co-shared articles than in articles from the same reliable sources that are not co-shared.}
\label{fig:method_example}
\end{figure*}

For example, consider an article (\url{washingtonpost.com/politics/2022/11/23/vaccinated-people-now-make-up-majority-covid-deaths/}) published by The Washington Post (\abr{WaPo}) with the headline ``Vaccinated people now make up a majority of covid deaths'' (which has since been changed to ``Covid is no longer mainly a pandemic of the unvaccinated. Here’s why.''). 
This article was published by a mainstream source and the headline is, strictly speaking, true. 
It was disproportionately shared by users in our data who \textit{also} shared anti-vaccine-related articles published by unreliable sources (Fig. \ref{fig:method_example}b). 
Without important context --- that since the vast majority of the country was vaccinated, the count of COVID-related deaths would be higher among the vaccinated even if the infection \textit{rate} remained much lower --- one can see how this content would be useful for a user who wished to promote the false narrative that the COVID-19 vaccine is ineffective or actively harmful. 
This narrative is popular in COVID-19-related misinformation content \cite{sharma2022covid}, and is consistent with work showing how vaccine skepticism can be promoted through misleading content rather than direct falsehoods \cite{allen2023quantifying}.
As this \added{illustrative} example suggests, those who wish to promote potentially misleading narratives such as these can strategically repurpose factually correct information in order to do so, when such information is available.

We test our expectations by deploying a tool for automatically extracting narrative structures from English text that are represented using narrative strings or labels. 
This provides us with narratives present in fake news articles as well as claims fact-checked to be false (Fig. \ref{fig:method_example}a). 
Our experiments across a variety of collections of potentially misleading narratives and types of reliable outlets support our primary hypothesis (the testing procedure is visualized in Fig. \ref{fig:method_example}c). 
The difference in the presence of potentially misleading narrative strings between co-shared and ``control'' articles (those published by the same set of mainstream news outlets but not co-shared with articles from unreliable sources) remains significant when accounting for partisan differences in the Twitter audiences of the two groups (\emph{SI Appendix}, section S2).  

Our approach incorporates and systematizes conceptual frameworks in misinformation research that highlight users' own roles in advancing false and misleading claims \cite{prochaska2023mobilizing, starbird_influence_2023}.
Understanding the dynamics that advance these broader claims, and not just specific pieces of incorrect information, is critical --- strictly true information repurposed to advance misleading claims can have harmful real-world consequences (such as reduced vaccine uptake) \cite{allen2023quantifying}.
Methodologically, we provide a scalable and generalizable way of identifying misinformation that would not be captured by domain-level classification, complementing recent text-based approaches \cite{wirtschafter2024detecting}.

Finally, our findings have important implications for responsible journalistic practice.
When vetting news stories (including the headlines) and their framing, it is important to consider not just whether the information contained in a headline or story is strictly true, but also whether it is likely to be \textit{used} in ways that are more broadly misinformative.
We further highlight the possible consequences of framing choices in mainstream news and their potential to be repurposed for misinformation through \replaced{two}{a} qualitative case \replaced{studies}{study}, contextualizing how our framework can support journalism in practice. 
In particular, our qualitative case studies highlight how particular mainstream news articles co-shared with fake news can be repurposed to spread potentially misleading narratives: we uncover different types of co-shared mainstream articles in terms of how they can be repurposed. These case studies cover two different topics: vaccine-related misinformation (of the form that imply vaccinated people spreading the disease or being the key node of transmission, especially in the context of COVID-19) and voter-fraud related misinformation (allegations of voter fraud in the 2020 U.S. election, particularly related to mail-in ballots). Both these topics of misinformation were prominent on Twitter and online social media in 2020 \cite{rosenberg2020twitter,ebermann2022conspiracy,berlinski2023effects}.

\begin{figure}[!t]
\centering
\includegraphics[width=0.99\textwidth]{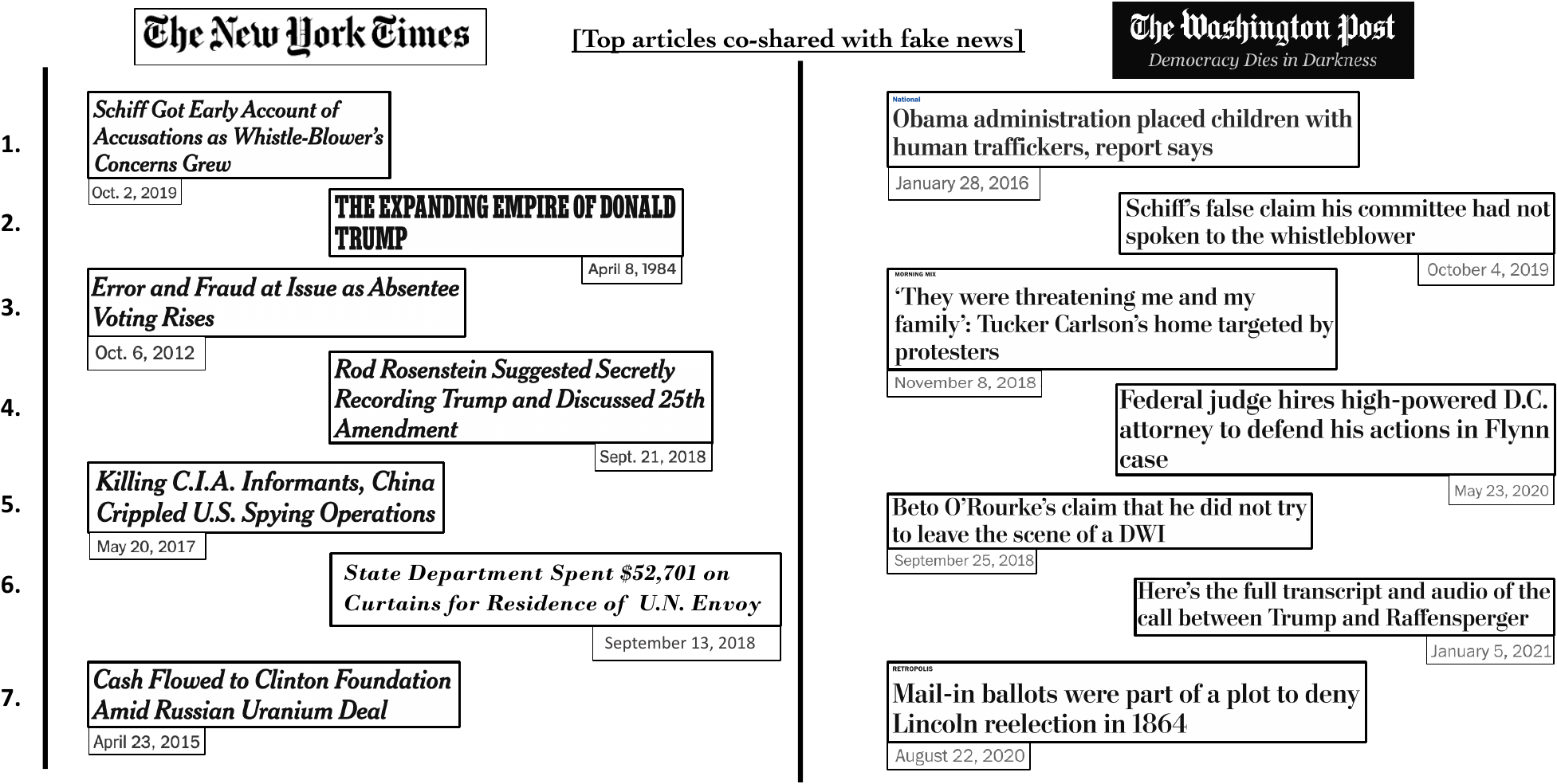}
\caption{\textbf{The top articles published by The New York Times (\abr{nyt}) and The Washington Post (\abr{WaPo}) in the set of articles disproportionately co-shared with fake news.} This is based on the co-sharing likelihood derived from Twitter sharing behavior (specifically, based on an aggregated co-sharing likelihood score for each outward edge from a particular article or \abr{url}). As motivation for our hypothesis in this work, consider the third \abr{nyt} article and the seventh \abr{WaPo} article shown in the figure about mail-in ballots and absentee voting (which also pop up in our second qualitative case study). These articles are popular in the same Twitter circles sharing stories published by fake news outlets containing narratives (extracted in this work) that include `mail ballot raise risk fraud', `lot people cheat mail', `mail balloting increase incident fraud', `paper mail balloting mean dramatic increase fraud', and `republican complain potential mail absentee ballot fraud'. While only headlines are shown here for illustrative purposes, our investigation also used the entire text of the news article. In this figure, we are only showing the top articles that are not opinion pieces, but we note that especially for \abr{WaPo}, the top co-shared articles set does include opinion pieces.}
\label{fig:top_nyt_wapo_coshared}
\end{figure}
\begin{figure}[!t]
\centering
\includegraphics[width=.9\linewidth]{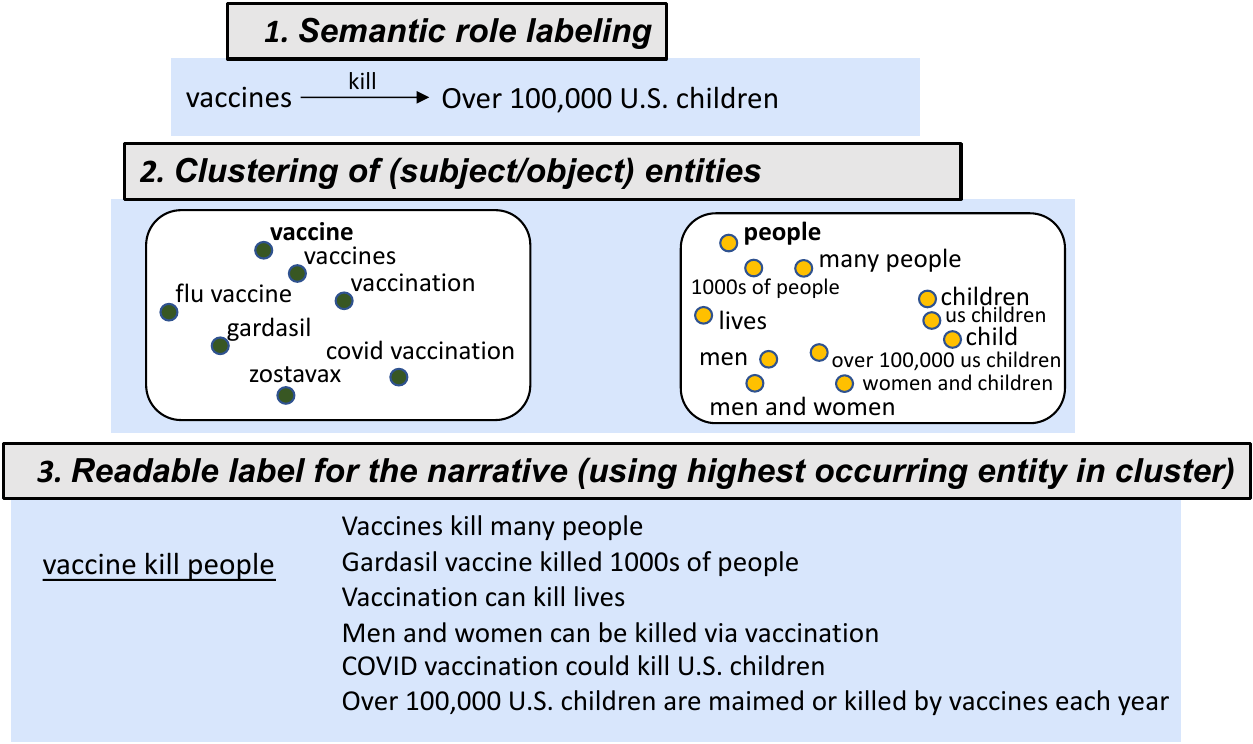}
\caption{\textbf{Brief summary of the process of extracting narratives.} This uses the method created in prior work~\cite{ash2023relatio}, and the visual is for the following sentence present in a fake news article in our dataset: ``Over 100,000 U.S. children are maimed or killed by vaccines each year.''}
\label{fig:narr_extraction}
\end{figure}

Throughout this paper, we present examples of various kinds from our dataset to help develop a theoretical intuition for our research design and findings. 
Fig. \ref{fig:method_example}, as mentioned above, presents a visual overview of our entire approach, using an illustrative example. 
Fig. \ref{fig:top_nyt_wapo_coshared} provides the top co-shared articles from The New York Times and The Washington Post. 
Fig. S1 in the \emph{SI Appendix} similarly presents the top co-shared articles from CNN and Fox News. 
Fig. \ref{fig:narr_extraction} provides a summary of the narrative extraction method using an example from our dataset. 
Table \ref{tab:misinfo_narr_examples} provides the top 3 most frequent extracted narratives that are present in misinformation content as well as co-shared mainstream articles from reliable outlets, but are absent in articles from the same reliable outlets that are not co-shared with ``fake'' news. 
Our case studies are also a collection of various kinds of examples touching all parts of our dataset: we provide examples of many relevant narratives, selected mainstream news articles along with the fake news articles they were co-shared with, and examples of publicly available tweets (in our dataset) by Twitter users with a history of sharing relevant fake news, sharing selected co-shared mainstream articles --- these examples provide valuable qualitative insights into \emph{how} mainstream news articles might be getting used for misleading purposes. 
These details and examples emergent in our case studies are provided in our \emph{SI Appendix}: sections S3 and S4, and Tables S5-S25. 
The case studies also highlight a few of the mainstream articles that gain disproportionate popularity among fake news circles on Twitter in our dataset. 
Tables S46 to S66 present all the narrative structures automatically extracted from claims fact-checked as false.

In this paper, we use the terms \emph{domains}, \emph{sources}, and \emph{outlets} interchangeably to refer to the organizations that produce political information. 
To facilitate a clearer understanding of our methods and findings, we provide definitions and brief contextual background for the terminology used throughout this paper below:

\emph{Narratives}: These are sets of 
relationships or structures in language that describe how entities act on each other \cite{sloman2015causality}, characterized by a base component of the following format: $\text{AGENT ENTITY} \xrightarrow{\text{VERB}} \text{PATIENT ENTITY}$ \cite{ash2023relatio}. 
Some examples are shown in Fig. \ref{fig:method_example}a and Table \ref{tab:misinfo_narr_examples}; the extraction process is summarized in Fig. \ref{fig:narr_extraction}. 
This is a specific instantiation using some of the components that have been found to be crucial for characterizing narratives in text across the broad and varied literature on narratives (as well as timing, motivations, etc. \cite{piper2021narrative}). 

\emph{Misinformation}: This broadly refers to false information.
This often includes content published by outlets categorized as unreliable and specific claims fact-checked as false. 
The former may also be referred to more specifically as ``fake'' news, which is typically considered to be a subset of misinformation. 
Misinformation is also distinguished from \textit{dis}information in that the latter involves an organized attempt to deceive. 

\emph{Fake news outlets} and \emph{fake news}: Fake news outlets are sources that produce news via processes that do not conform to standard industry practices (including editorial norms and processes for vetting information) \cite{lazer2018science,grinberg2019fake}.
The definition we use here is closely related to the usage of `the deliberate creation of pseudojournalistic disinformation' in prior work \cite{egelhofer2019fake}.  

We classify outlets known to regularly publish false content as fake news sources, and articles published by those sources as fake news, using annotations from \newsguard{} \cite{newsguard}. 
We aim for precision by considering `regularly publishing false content' as our criterion out of all the factors that determine the trustworthiness score of an outlet in \newsguard{} ratings. This helps us highlight the drawbacks of conceptualizing problems with the digital information ecosystem as that of `false information' or fake news content --- a framework that often foregoes closer examination of information coming from sources deemed reliable. Complete details about the use of \newsguard{} annotations in our work, description of \newsguard{} and their methodology, and the rationale for using these annotations are provided in the \emph{Methods} section. 

\emph{Reliable outlets}: These are all news outlets annotated for their quality as information sources by journalists at \newsguard{} and \emph{not} deemed as fake news outlets according to our selection criteria above. 
Since this is quite broad, we also show our primary results with narrower criteria such as only considering sources marked as `trustworthy' by  \newsguard{}. as well as partisanship-based subsets (details provided in the \emph{Methods} section). 


\emph{Mainstream media or news (outlets)}: Within reliable outlets, we define mainstream outlets as those that \emph{publish predominantly political content} and are \emph{popular}. 
We borrow Twitter-derived domain-level ratings as well as site-visiting statistics from passive metering created in prior work \cite{greencuration,guess2021almost}. 
Details on how the two criteria are measured and applied are presented in the \emph{Methods} section. 

\emph{Trustworthy outlets}: These outlets are a particular subset of the set of mainstream media outlets defined above: ones that are annotated as `trustworthy' in \newsguard{} ratings (based on criteria that have to do with journalistic quality and standards of transparency and credibility). We note again that `regularly publishing false content' --- our criterion for deciding the outlets that are considered as fake news outlets in this study --- is one of several criteria that determine trustworthiness scores in \newsguard{} ratings (details provided in the \emph{Methods} section). 


\emph{Potentially misleading narratives}: These are narratives present in false claims or frequently recurring across content published by fake news sources. 
These extracted narratives, such as `vaccines harm people', are not by construction true or false, and cannot be independently judged for truthfulness.
They are simply prevalent in misinformation content. 
However, potentially misleading narratives need not be limited to content present in fake news outlets.

Our approach for automated narrative extraction is detailed in the \emph{Methods} section (also summarized in Fig. \ref{fig:narr_extraction}, examples in Table \ref{tab:misinfo_narr_examples}). 
For the sake of being comprehensive, we include results where \emph{all} structures extracted from fake news content are also used to test our hypothesis, not just recurring ones. 
However, we define potentially misleading narratives as those narratives that either occur repeatedly in fake news content (where each instance, in trying to mirror mainstream news, contains a detailed story spanning multiple sentences), or the ones that occur in false claims (where each instance is a fact-checked false claim consisting of a short sentence). 
To enable greater transparency about automatically extracted narrative labels (or string representations), we provide all narratives extracted from false claims (used to test our hypothesis) in \emph{SI Appendix}, section S6 and Tables S46 to S66. 



Our initial dataset consists of tweets collected from a panel of over $1.6$ million U.S. Twitter users matched to their voter registration records via a commercial U.S. voter file (described in prior work \cite{hughes2021using}). 
This panel slightly over-represents white and female Twitter users, but is otherwise demographically similar to samples of Twitter users collected via high-quality surveys \cite{hughes2021using,shugars2021pandemics}. 
We retain English tweets that share news articles (published by English-language information sources or outlets per \newsguard{}) from May 1 2018 to November 14 2021. 
For each tweet, we have the user \abr{id} of the author, and the news \abr{url} being shared in the tweet. 
We keep \abr{url}s that were shared by at least $20$ different users (to control for sparsity effects), resulting in ${\sim}420,000$ different \abr{url}s, published by ${\sim}2400$ domains, and shared by ${\sim}450,000$ different users. 

The methodological approach used in this work is visually illustrated in Fig. \ref{fig:method_example}. 
Our process for finding co-shared mainstream articles is visualized in Fig. \ref{fig:method_example}b. 
We construct a weighted graph with individual articles or \abr{url}s as nodes. 
Edges between a fake news \abr{url} (nodes shown on the left-hand side of Fig. \ref{fig:method_example}b) and a reliable news \abr{url} (nodes shown on the right-hand side of Fig. \ref{fig:method_example}b) are weighted by the number of Twitter users in our dataset that shared \emph{both} the incident node \abr{url}s at least once. 
We apply a graph-pruning algorithm \cite{dianati2016unwinding} to assign each edge a co-sharing score based on the likelihood of both incident node \abr{url}s getting shared by the same user, controlling for the individual popularity of each of those \abr{url}s in isolation. 
We consider articles published by mainstream outlets that fall into this top $1\%$ set of articles as \emph{the group of mainstream news articles that are co-shared with fake news}. 
Fig. \ref{fig:top_nyt_wapo_coshared} displays examples of top co-shared articles published by \textit{The New York Times} and \textit{The Washington Post}, and similarly Fig. S1 in \emph{SI Appendix} displays top co-shared articles published by CNN and Fox News. 
Stories published by the same set of mainstream outlets that are \emph{not} in the top $5\%$ of co-sharing likelihood scores form our \emph{control group}. 

We operationalize narratives using methods and an open-source package developed in prior work \cite{ash2023relatio}, available at \url{github.com/relatio-nlp/relatio}.  
Fig. \ref{fig:narr_extraction} briefly visualizes the process of extracting the narrative structure through an illustrative example. 
Extracted narratives are simply represented as strings, or what we call narrative labels or narrative strings (`vaccine cause shingles'; `vaccine cause people develop shingles'). 
Two settings are used to extract narratives at two levels of granularity from the text data (following \cite{ash2023relatio}): \emph{low-dimensional narratives}, allowing fewer, large-sized clusters or latent entities that result in `broader' narratives, and \emph{high-dimensional narratives}, allowing a higher number of smaller-sized clusters that result in more specific or `fine-grained' narratives. 

To obtain sets of potentially misleading narratives, we consider two sources of data. 
The first consists of the aforementioned fake news articles (${\sim}24k$ \abr{url}s) shared in our Twitter dataset, where we consider the headline and story for each article as separate text instances, since many (at least 6 out of 10) people on social media only read the headline of an article \cite{gabielkov2016social}; see also: \url{americanpressinstitute.org/publications/reports/survey-research/how-americans-get-news/} and \url{washingtonpost.com/news/the-fix/wp/2014/03/19/americans-read-headlines-and-not-much-else/}. 
 Headlines and story texts are deduplicated and processed to remove information like domain name mentions and headline text within the story text. 
This results in ${\sim}30k$ unique texts derived from fake news articles. 
Potentially misleading narratives in fake news content are defined as recurring narratives: the top $1\%$ in terms of frequency of occurrence (therefore, these narratives occur across several texts in our dataset). We also experiment with all the narratives found in fake news content (Table \ref{tab:misinfo_narr_libs}). 
The second source consists of texts of claims that have been fact-checked as false (${\sim}24k$ unique texts or claims). 
For fact-checked false claims, we collect and combine various publicly available datasets (explicated in \emph{Methods} section), and use the claims that are labeled as false (or an equivalent label such as `not true'). 
Since these are statements, usually consisting of a single sentence fact-checked as false, we consider all narratives extracted from this data as our set of potentially misleading narratives. 
This results in three libraries of potentially misleading narratives, each containing low- and high-dimensional narratives (Table \ref{tab:misinfo_narr_libs}). 
Table \ref{tab:misinfo_narr_examples} presents examples of potentially misleading narratives.

The \emph{Methods} section provides details on the construction of the underlying Twitter panel dataset (including matching between Twitter and U.S. voter data), data collection, sampling of the Twitter dataset, demographic information for the Twitter dataset, process for obtaining news article content, graph-pruning method for obtaining co-shared mainstream articles and constructing the control group, the narrative extraction method and rationales for all the various choices made throughout the method design (such as the choice of date range and thresholds for selecting co-shared articles). 



\section*{\added{Results}}

The co-shared and control groups of mainstream news articles and different sets of potentially misleading narratives enable us to test our hypothesis: \textbf{mainstream news stories that are disproportionately co-shared with fake news are significantly more likely to contain potentially misleading narratives \added{(narratives that are prevalent in misinformation content)}.} 
Put another way, we test whether potentially misleading narratives occur more frequently in the set of co-shared articles than in the set of control articles published by the same mainstream outlets. 
Our testing mechanism is visually \replaced{illustrated}{explained} in \cref{fig:method_example}c. 
We compare the counts or the presence of a set of potentially misleading narratives in co-shared articles with the control set using a one-sided Wilcoxon-signed rank test, a non-parametric significance test for comparing paired or dependent data \cite{wilcoxon1992individual,woolson2007wilcoxon} (the procedure for obtaining counts of relevant narratives in particular articles is explained in the \emph{Methods} section). 
\added{Since the \emph{same} set of narrative labels are being compared for their occurrence in two groups, our scenario fits the paired data testing scenario, and a non-parametric test like Wilcoxon allows us to proceed without having to assume normality of the underlying distributions. 
While we use a one-sided test here since our hypothesis is directional (testing that the presence of particular narratives in one group is higher than the other), using a two-sided Wilcoxon-signed rank test replicates all of our main results (see Section S9 and Table S45 in the \emph{SI Appendix}).}
Any occurrences of ties and zeros were removed before this test is performed \cite{pratt1959remarks}, resulting in the final counts or $n$ shown in \cref{tab:misinfo_results_sig_testing}. 

This process is repeated for all three of our potentially misleading narrative libraries (see \cref{tab:misinfo_narr_libs}), with the test performed separately for high- and low-dimensional narratives. 
To control for distributional shifts in the number of articles published by reliable domains, we also perform the same test with articles in both groups limited to those published by trustworthy outlets, as well as articles in both groups limited to those published by liberal outlets alone (the process for obtaining liberal-conservative classification of outlets is described in the \emph{Methods} section). 
This check accounts for the fact that even when the two groups of mainstream articles are published by the same set of domains or outlets, the \emph{number} of articles published by each domain can constitute different distributions. We show the distribution of per-outlet-article-counts for the two groups of articles from reliable outlets in the \emph{SI Appendix} (section S1 and Fig. S2). 
Since there is a conservative lean when considering all outlets or even just trustworthy outlets, we repeat our experiments with articles published by liberal outlets alone. 

The results across all these settings, shown in \cref{tab:misinfo_results_sig_testing}, support our hypothesis: articles from mainstream outlets that are co-shared with fake news contain \replaced{narratives extracted from misinformation content}{misinformation narratives} at a significantly higher rate.  
This suggests that potentially misleading narratives may allow for a strategic repurposing of mainstream news in fake-news-sharing circles on Twitter, something our data and method can be used to investigate as demonstrated by our case stud\replaced{ies}{y} (described \deleted{briefly} in the next section). 
The significance testing procedure is detailed in \replaced{the \emph{Methods} section}{\emph{Materials and Methods}}. 

We also estimate the magnitude of the difference in co-shared and control sets across the full set of mainstream news articles\added{, while emphasizing that it is the difference between the two groups that is key to our hypothesis}. 
After combining all recurring narratives in fake news and all narratives in false fact-checked claims as our set of potentially misleading narratives, we count the potentially misleading narratives present in each article. 
Roughly one ($0.94$) potentially misleading narrative is present in a co-shared mainstream article on average, whereas for the control set, a potentially misleading narrative is present in only every other article on average ($0.58$ potentially misleading narrative per article). 
When dividing by the total number of extracted narratives present in each article, so as to compare ratios instead of counts, we find that, on average, $2.2\%$ of all narratives in a co-shared article are present in the set of all narratives extracted from misinformation content (recurring across fake news or present in fact-checked false claims), whereas $1.3\%$ of all narratives in a control article are present in the same set of all narratives extracted from misinformation content (recurring across fake news or present in fact-checked false claims). This comparison (average and standard deviation) is also noted in Table \ref{tab:misinfo_quant_presence_coshared_control}. 

Since we have a large number of texts (more than $50,000$ unique texts across fake news articles and headlines, and false fact-checked statements), we are ultimately working with a set of over $27,000$ different narrative labels or narratives when checking the number of potentially misleading narratives present in a particular mainstream news article. 
Although some occurrence of potentially misleading narratives is, therefore, expected in mainstream articles due to factors such as shared language in news reports, our results in Table \ref{tab:misinfo_results_sig_testing} show that their occurrence is significantly higher in mainstream articles co-shared with fake news articles. 
Put another way, the value of $0.58$ or $1.3\%$ represents the \emph{noise} --- the expected baseline overlap that occurs with mainstream texts due to the methodology of narrative extraction itself --- and the `0.94' or $2.2\%$ value for co-shared texts represents the significantly greater \emph{signal}. 

\paragraph{\added{Partisanship as a differing factor between co-shared and control groups.}}
It is possible that these results \replaced{depend on}{are driven by} the co-shared and control articles differing along another dimension other than their propensity to include potentially misleading narratives --- namely, partisanship.
In \emph{SI Appendix} (section S2), we conduct a series of robustness checks to test for the possibility that the relationships we observe are instead attributable to partisan curation.
We find that while right-leaning articles are more likely to contain narratives present in misinformation content, articles in the co-shared group remain significantly more likely to contain those potentially misleading narratives even when controlling for audience partisanship (with and without domain-level fixed effects).
Articles in the co-shared group have 
roughly $1.24$ times the odds of the control group of containing potentially misleading narratives (Tables S2 and S3, \emph{SI Appendix}).

\section*{Case Stud\replaced{ies}{y}}

Our quantitative results show that narratives are significantly associated with co-sharing of certain mainstream articles in the \emph{aggregate}. 
In addition, our data and methods can help contextualize the relationship between information and narratives, and how particular mainstream articles co-shared with fake news can be used or repurposed to spread potentially misleading narratives. 
To this end, we conduct two case studies, covering two different topic areas: vaccine-related misinformation and voter-fraud related misinformation, both prominent on Twitter and online social media in 2020 \cite{rosenberg2020twitter,ebermann2022conspiracy,berlinski2023effects}. 
These case studies highlight how our work can contribute to the understanding of specific narratives in the media landscape, as well as enable journalists and scholars to help investigate potential cases of synergistic support for potentially misleading narratives in mainstream news.

\subsection*{\added{Case Study 1: Vaccination}}

In our manual examination of narrative labels (and narratives) present in co-shared articles published by liberal mainstream outlets, we find multiple labels of the form `vaccinated people spread disease', `vaccinated people spread virus', and `vaccinated people spread delta variant'. 
We present details of our manual filtering approach in \emph{SI Appendix}, section S3, and all such narrative labels obtained along with sentences containing them in \emph{SI Appendix}, Tables S6 to S8. 
Here we examine the nine mainstream articles that are \emph{a)} published by \emph{liberal} mainstream outlets; \emph{b)} co-shared with fake news; and \emph{c)} contain the aforementioned identified type of narrative. 
The article \abr{url}s, their headlines, relevant story snippets, and relevant narrative labels they contain are provided in \emph{SI Appendix}, Tables S9 to S11. 
These nine articles can be grouped qualitatively into three types in terms of their propensity to be used for misleading purposes: 

\begin{itemize}
    \item \textbf{Type 1: Clickbait headlines.} We find that three mainstream articles contain at least a mild form of clickbait or misleading headlines, via trivialization or simplification that could provide easy support for misleading anti-vaccination narratives and beliefs. Headlines carry particular significance in their much wider reach and readership \cite{dor2003newspaper,gabielkov2016social} as well as their influence on how the article, if read, is interpreted \cite{ecker2014effects}. When written for the primary purpose of generating clicks rather than to inform, headlines can omit context in ways that sensationalize by presenting the news item as ``more interesting, extraordinary and relevant than might be the case'' \cite{molek2013towards}, and such clickbait headlines can be actively misleading \cite{chen2015misleading}. 
    A recent study suggests that the pragmatic implications of news headlines influence the readers' reactions, and reasoning about these implications of headlines is important to combat misinformation \cite{gabriel2022misinfo}. 

    For example, the headline ``Top health expert says vaccinated people are spreading delta variant'' (published by The Hill
    ; first row in \emph{SI Appendix}, Table S9) can be used to ascribe a causal or unique role to vaccinated people in the spread of the variant as opposed to unvaccinated people (who are not implicated via omission). 
    The article itself is careful not to suggest this, noting that ``...public health experts are encouraging further vaccinations to help curb transmission. Some experts, \emph{however}, warn that vaccinated individuals \emph{may still} be capable of contracting and transmitting COVID-19'' (emphasis added). The addition of uncertainty is in direct contrast to the headline. Further, this snippet clarifies that rather than having a unique or causal role in transmissions, vaccines do not completely eliminate transmissions. This case is related to that of incongruent headlines \cite{chesney-etal-2017-incongruent}, when statement(s) in the headline are not supported in the article's content, thereby misleading readers. 
    This case also features an example of what linguists would call a quantity implicature \cite{grice1975logic}, in which an ambiguous statement (e.g. ``Q: Who came to the party? A: Amy and Bonnie'') is interpreted to exclude other possible alternatives (implying that only Amy and Bonnie came to the party, although the answer is true if additional people also came). 
    Here, the phrasing of the headline permits and arguably encourages restrictive interpretations in which vaccinated people are spreading the delta variant at equal or higher rates than unvaccinated people, supporting a narrative in which vaccines are ineffective or actively harmful --- despite the article itself making clear this is not the case. 
    
    The use of the headline to support anti-vaccine worldviews is further illustrated in the text of the tweets sharing these articles (\emph{SI Appendix}, Table S13), such as `Vaccines that target the spike protein are causing it to mutate. Virology 101. \dots{}' and `\dots{} THAT IS LITERALLY THE HEADLINE. I wonder if President Biden will consider me a ``murderer'' for sharing.'  

    \item \textbf{Type 2: Older news articles in new contexts.} These are articles published well before the period we analyze (May 2018 to November 2021) that report on particular vaccine failures in the past, reintroduced by Twitter users to make claims about contemporary vaccines. As shown in \emph{SI Appendix}, Table S10, we find two such articles, one from mid-2017 by NPR 
    (headline: ``Mutant Strains Of Polio Vaccine Now Cause More Paralysis Than Wild Polio''), and another from 2015 by PBS (
    headline: ``This chicken vaccine makes its virus more dangerous''). 
    
    Sharing older news articles in a new context can be used to mislead or as a disinformation tactic; news organizations have taken steps to highlight their publication date \cite{rogers2019latest}. In 2020, Facebook rolled out a feature to notify users when they are sharing articles more than $90$ days old (see \url{dailymail.co.uk/sciencetech/article-8460273/Facebook-tell-users-share-old-news-article.html} and \url{forbes.com/sites/alisondurkee/2020/06/26/new-facebook-policy-warns-users-before-they-share-old-articles/?sh=3b0eb9fb3f34}). 
    Tweets by users promoting vaccine misinformation sharing these articles (\emph{SI Appendix}, Table S13) can further help understand such repurposing of prior, legitimate reporting. For example, the aforementioned polio vaccine article by NPR was shared with comments such as  `None for me, thanks. I'll take my polio natural.' and  `2 year old article is still relevant., Nope. Vaccines have been causing VARIANTS for a LONG time. All you folks that got VACCINATED are the ones who have ``caused this''. \#TheVaccinesAreTheVariants'. 

    \item \textbf{Type 3: Other.} These are the remaining articles that do not seem to be obviously usable for misleading purposes. 
    
\end{itemize}

We show a sample of the vaccine-related fake news articles co-shared with the aforementioned mainstream articles in \emph{SI Appendix}, Table S12. 
This provides exposure to the interactions between fake news and co-shared mainstream articles. 
A random sample of tweets from users promoting vaccine-related fake news who also shared these articles from liberal-leaning mainstream outlets are provided in \emph{SI Appendix}, Table S13, while \emph{all} the unique tweet text from users who shared vaccine-related fake news and both types of mainstream articles are also presented in \emph{SI Appendix}, section S3. 
These tweets illustrate \emph{how} these `real' news articles might be getting used to support potentially misleading narratives.

Finally, we take a closer look at the three mainstream articles (\emph{SI Appendix}, Table S9) with clickbait headlines, and quantify their impact in terms of their disproportionate circulation among users promoting vaccine misinformation (by having a history of sharing vaccine-related fake news articles). 
We find that these particular articles by the three mainstream outlets (The Hill, Business Insider, and \textit{The Washington Post}) are some of their most popular vaccine-related articles among these users, relative to all the vaccine reporting done by those outlets (per the \abr{url}s shared in our Twitter dataset between May 2018 and November 2021). 
For example, out of all the $444$ vaccine-related articles published by \textit{The Washington Post} in our final dataset, the particular co-shared article with the headline ``CDC study shows three-fourths of people infected in Massachusetts coronavirus outbreak were vaccinated but few required hospitalization'' was the $2^{nd}$ highest in terms of shares amongst the subset of Twitter users known to share vaccine-related fake news (\url{washingtonpost.com/health/2021/07/30/provincetown-covid-outbreak-vaccinated}). 
We also find that these articles were disproportionately shared by uses who also shared vaccine-related fake news, relative to a comparison group of users who shared reliable vaccine information. 

We provide all relevant sets of articles and tweet text along with details of various procedures involved in this case study in \emph{SI Appendix}, section S3, including the process of obtaining the vaccine-related fake news articles co-shared with mainstream articles of interest and significance tests that highlight the disproportionate uptake of these mainstream articles with clickbait headlines among users who share vaccine-related fake news.


%

\subsection*{\added{Case Study 2: Voter Fraud Allegations in the 2020 U.S. Elections}}

\added{One of the most widespread mis- or disinformation campaigns of 2020 in the U.S. was allegations of voter fraud in the 2020 U.S. election, which can negatively impact people's trust and confidence in the election process \cite{ebermann2022conspiracy,berlinski2023effects}. Unsurprisingly, we find many narrative labels capturing claims of voter fraud, especially focusing on the use of mail-in ballots since mail-in based voting was significantly expanded in the 2020 elections in the wake of the COVID-19 pandemic and social distancing recommendations. Much like case study 1 above, we conduct an examination of co-shared mainstream articles published by liberal outlets containing narratives that could capture voter fraud claims, and analyze the possibility of some mainstream news articles providing potential ammunition to users who have a history of sharing fake news. }

We compile relevant extracted narratives by first filtering all narrative labels containing `fraud' and `mail', `faulty' and `ballot', and other keywords that emerge upon manual examination such as `behalf' --- we also filter out irrelevant narratives that directly suggest a focus on `allege' and `speculate' instead of asserting a claim. We manually select similar relevant and coherent narratives together under an umbrella for this case study, allowing for a lack of precision to try to capture maximal cases of such narratives and potential articles to examine qualitatively. We provide details of our process, all $150$ narratives that emerged via keyword search, and all $49$ relevant narratives that emerge upon manual examination along with sentences containing those selected narratives in the \emph{SI Appendix}, section S4. These narratives are then used to find relevant mainstream co-shared articles. 

\added{Specifically, we manually examine all mainstream articles that are \emph{a)} published by \emph{liberal} mainstream outlets; \emph{b)} co-shared with fake news; and \emph{c)} contain at least one of the selected narratives that pertain to voter fraud. 
For all these of mainstream articles, the article \abr{url}s, their headlines, relevant story snippets, and relevant narrative labels they contain are provided in \emph{SI Appendix}, Tables S19 to S22. After examining these articles, we find $11$ instances of articles that could potentially be used to support a narrative or worldview of widespread voter fraud. These $11$ articles can be grouped qualitatively into three types, in terms of how they might be useful for misleading purposes: }

\begin{itemize}
    \item \added{\textbf{Type 1: Mainstream articles easy to use for promoting mail-in vote based fraud conspiracy theory.} These articles are interesting cases where the content, including the headline, provides easy use or re-use for promoting the voter fraud narrative in 2020 in the U.S.: \emph{i)}} \url{nytimes.com/2012/10/07/us/politics/as-more-vote-by-mail-faulty-ballots-could-impact-elections.html} (with the headline ``As More Vote by Mail, Faulty Ballots Could Impact Elections''); \added{\emph{ii)}} \url{washingtonpost.com/history/2020/08/22/mail-in-voting-civil-war-election-conspiracy-lincoln}\added{ (with the headline ``Mail-in ballots were part of a plot to deny Lincoln reelection in 1864'');} \added{\emph{iii)}} \url{thehill.com/opinion/campaign/506331-vote-by-mail-would-create-chaos-and-distrust-in-november}\added{ (with the headline ``Vote-by-mail would create chaos and distrust in November''). }

    The report from The New York Times was published in 2012, but given its suggestion of vote-by-mail increasing chances of faulty ballots and having an impact on elections, it provides a perfect mainstream story to support the voter fraud conspiracy in 2020. This is also an example of temporal repurporsing or using an article from a different time and context (as discussed about `type 2' articles in case study 1), to promote a certain narrative in a different time and context. The article's story content clarifies that the number and impact of mail-in ballots is not as clear cut, but given the strong push in 2020 toward a belief in fraud and questioning the election outcome even before it occurred, even just the headline lends itself to the conspiracy and can be \emph{easily} used towards this end (indeed, as we establish below, since this article from The New York Times is disproportionaly popular among voter-fraud or mail-in ballot-related misinformation sharers on Twitter in our dataset, compared to all reporting on voter fraud from The New York Times in our dataset). We can see strategic use of this article, including emphasis on the domain name (The New York Times) to take advantage of the outlet's reputation, in tweets sharing this article by users who have shared voter-fraud or mail-in ballot related fake news; for example, the tweet ``Even The NY Times agreed that vote by mail ballots could impact the 2012 election. So why can’t they impact the election in 2020? Well...'' and the tweet ``FLASHBACK NEW YORK TIMES (2012): As More Vote by Mail, Faulty Ballots Could Impact Elections hat tip:''. 

    The Washington Post article similarly can be easily and directly used to promote voter fraud conspiracy theories, especially those centered on mail-in ballots. The headline itself --- ``Mail-in ballots were part of a plot to deny Lincoln reelection in 1864'' --- clearly states mail-in ballots were used to try to influence an election outcome. Perhaps alarmingly, this article was published in August 2020, when misinformation about mail-in ballot-based fraud and its exaggerated impacts on election outcomes had plenty of momentum, especially on social media. Furthermore, this article and headline focuses on a deliberate, intentional plot to impact an election, and intentional fraud and 'stealing the election' is at the heart of the conspiracy in 2020. While the subject of this story is an election that occurred in 1864, long before 2020, it provides an instance of a story from a mainstream outlet that can be easily used to prop up misinformation around electoral fraud in 2020. In fact, given the timing of this article and a focus on the intent to impact an election, this might be the most useful mainstream article for supporting the conspiracy that we have uncovered in our data. 

    \added{Finally, The Hill article is an opinion piece, making a negative case for vote-by-mail, conceding the lack of trust and ``chaos'' it would create. It also claims mail-in voting is a cause of fraud, surfacing select research while ignoring other research. Since a mainstream outlet is making the case \emph{against} vote-by-mail, it can be directly and easily used by people interested in using expanded mail-in voting to suggest electoral fraud.}

    \item \added{\textbf{Type 2: Mainstream articles reporting on actual cases of voter fraud, or allegations with a concrete basis.} The thrust behind the voter fraud conspiracy theory for the 2020 election is not that individual-level fraud or mistakes never occurs, but that such fraud has occurred or will occur in a `widespread' manner and at a high enough rate so as to impact the election, which is the false claim or belief. Although the claim asserts a large magnitude, people making this false claim can use individual cases of actual fraud, or concretely alleged fraud, as `proof' of the alleged widespread nature of the phenomenon. To this end, mainstream reports on individual cases of fraud or alleged fraud are useful for misinformation spreaders as something concrete and believable to point to, reported by reputable mainstream outlets. } 

    \added{We find six such articles: }
    \begin{enumerate}
        \item \url{wsaz.com/2020/07/10/mail-carrier-in-west-virginia-pleads-guilty-to-attempted-election-fraud} (with the headline ``Mail carrier in West Virginia pleads guilty to attempted election fraud''); 
        \item \url{kron4.com/news/california/northern-california-woman-faces-felony-voter-fraud-charges-for-allegedly-voting-twice/1575491353} (with the headline ``Northern California woman faces felony voter fraud charges for allegedly voting twice''); 
        \item \url{nbcnews.com/news/us-news/republican-official-ohio-faces-charge-voting-twice-november-election-n1271985} (with the headline ``Republican official in Ohio faces charge for voting twice in November election''); 
        \item \url{politico.com/story/2018/12/04/north-carolina-elections-fraud-allegations-mark-harris-campaign-house-1045355} (with the headline ``GOP hit with election fraud claims after using issue as rallying cry''); 
        \item \url{businessinsider.com/voter-election-fraud-pennsylvania-charge-dead-mom-vote-trump-2020-12} (with the headline ``Officials finally found a case of a dead person voting, accusing a Republican of pretending to be his dead mom to vote for Trump''); 
        \item \url{thehill.com/homenews/531242-pennsylvania-trump-supporter-charged-with-voter-fraud} (with the headline ``Pennsylvania Trump supporter charged with voter fraud'').
    \end{enumerate}

    \added{Observing tweets by users who have a history of sharing voter fraud related fake news articles, we see the use of these individual cases to push for the widespread nature or push back against the refutation of conspiracy theory by mocking the idea that fraud cannot happen; for example, a user with a history of sharing voter fraud-related fake news shared the WSAZ News article (on a mail carrier in West Virginia pleading guilty to attempted election fraud) with the tweet: ``But but but this would never really happen!!! Mail carrier in West Virginia pleads guilty to attempted election fraud for attempting to alter mail-in ballot request forms.''}

    \item \textbf{Type 3: Mainstream articles containing mail-in ballot fraud claims from actors, without any pushback.} We also find two mainstream news articles that contain the false voter fraud claim (via some actor making the claim and getting quoted on that) without any pushback or fact check for the claim. In this way, the false claim is allowed to stand without providing any correction in a mainstream news report, which can potentially legitimize the false claim or be used for such legitimization by actors who believe the claim and want to spread it. The two such articles we find are: \url{thehill.com/homenews/house/492057-mccarthy-slams-democrats-on-funding-for-mail-in-balloting} (with the headline ``McCarthy slams Democrats on funding for mail-in balloting'') and \url{nbcnews.com/politics/meet-the-press/two-thirds-voters-back-vote-mail-november-2020-n1187976} (with the headline ``Two-thirds of voters back vote-by-mail in November 2020'').

\end{itemize}

We show a sample of the voter-fraud or mail-in ballot related fake news articles co-shared with the aforementioned mainstream articles in \emph{SI Appendix}, Tables S23-S24. 
This provides further examples of the interactions between fake news and co-shared mainstream articles. 
A random sample of tweets from users promoting voter-fraud or mail-in ballot related fake news who also shared these (type 1 or type 2) articles from liberal-leaning mainstream outlets are provided in \emph{SI Appendix}, Table S25. 
These tweets illustrate \emph{how} these `real' news articles might be getting used to support potentially misleading narratives.

Finally, just like in case study 1, we take a closer look at the three type-1 mainstream articles (\emph{SI Appendix}, Table S19) that can be very easily and directly used to promote the voter-fraud (particularly, mail-in vote-based fraud) conspiracy theory about the 2020 U.S. election, and quantify their impact in terms of their disproportionate circulation among users promoting voter-fraud or mail-in ballot related misinformation. 
We find that these particular articles by the three mainstream outlets (The New York Times, The Washington Post, and The Hill) are the respective outlet's most popular voter-fraud or mail-in ballot related articles among these users, relative to all the voter-fraud or mail-in ballot reporting done by those outlets (per the \abr{url}s shared in our Twitter dataset between May 2018 and November 2021). 
For example, out of all the $99$ voter-fraud or mail-in ballot related articles published by \textit{The Washington Post} in our final dataset, the particular co-shared article with the headline ``Mail-in ballots were part of a plot to deny Lincoln reelection in 1864'' was ranked the highest in terms of shares amongst the subset of Twitter users known to share voter-fraud or mail-in ballot related fake news. 
We also find that these articles were disproportionately shared by uses who also shared voter-fraud or mail-in ballot related fake news, relative to a control group of users who shared reliable voter-fraud or mail-in ballot related information. 

\added{We provide all relevant sets of articles and tweet text along with details of various procedures involved in this case study in \emph{SI Appendix}, section S4, including the process of obtaining the voter-fraud-related fake news articles co-shared with mainstream articles of interest and significance tests that highlight the disproportionate uptake of clickbait mainstream articles among users who share voter-fraud-related fake news.}

\section*{Discussion}

Consistent with our hypothesis, we find that mainstream articles that circulate among fake-news-sharing users on Twitter are significantly more likely to contain narratives that are prevalent in misinformation content, compared with mainstream articles from the same reliable outlets that are not co-shared with fake news. 
This effect is not fully attributable to partisan curation --- our findings are not entirely \replaced{dependent on}{driven by} potentially misleading narratives carrying a left- or right-leaning slant. 
Our finding suggests that users strategically repurpose mainstream news to develop and spread potentially misleading narratives on social media. 
Indeed, it is likely that users promoting misleading narratives find mainstream sources particularly attractive when they publish information that can be repurposed to fit those narratives \emph{precisely because of} these sources' credibility. 
This is consistent with related instances of partisan information being seen as more credible when it comes from unexpected sources \cite{baum2009shot}, and suggests a general framework in which users share information that they perceive to be \emph{useful} for advancing their broader worldviews.

Analyses based on our dataset show that the reach and potential audience for mainstream news articles co-shared with those fake news articles is \emph{more than twice} that of fake news articles.  
Specifically, on average, about $121$ users share the co-shared mainstream articles as compared to about $58$ who share a given fake news article.
Moreover, those who exclusively share fake news average roughly $23$ followers, while those who exclusively share co-shared mainstream articles averaged roughly $45$ followers (details are provided in \emph{SI Appendix}, section S7; the difference between the two groups is statistically significant). 
This nearly $2:1$ ratio is also observed when comparing the median numbers for (co-shared) mainstream and fake news article sets. 
This further underlines the need to study mainstream news sources and various articles published by reliable outlets, along with their interaction with the relatively niche presence of, and potential exposure to, fake news sources and articles. 
Potentially misleading narratives, by cross-cutting the type of outlet (fake news or mainstream), have a much larger platform than that of fake news alone. 

To the extent to which our finding about the role of mainstream news in online misinformation \added{networks on social media} is true, it has important implications for responsible journalistic practice: when vetting news stories and their framing, especially for the headlines, it is important to consider not just the raw information of the story itself but also the broader claims that the information could be used to support. 
Technological methods like the ones we have developed and used here may help go beyond checking a story's factuality and accuracy and contribute to minimizing the risk of a story being repurposed to mislead. 
Journalists committed to the rigor of their reporting and how their content might get used or misused can adjust their framing accordingly before publishing. 
This can help reduce the chances of mainstream news articles potentially legitimizing world views based on false or unsubstantiated claims. 

For scientific research on online misinformation, our work highlights the importance of considering the broader context in which a specific piece of information is being circulated, rather than solely relying on classifying information (to say nothing of sources) as strictly reliable or unreliable. 
This is connected to the idea of ``deep stories'', the larger underlying stories (or narratives) tying together various atomized daily stories occurring in people's lives \cite{hochschild2016strangers}, which can often be \emph{the} connective force behind people's beliefs regardless of plausibility and has been tied to fake news and disinformation in the U.S. \cite{polletta2019deep,prochaska2023mobilizing}. 
It is also important to examine the choices being made in mainstream news, such as the inclusion of existing misleading narratives. 
Indeed, the focus on the purely informational component of misinformation is insufficient to account for why misperceptions are widespread and persistent --- especially given the small market share of categorically ``fake" news outlets. 
Instead, a better framework for understanding the apparent problems with the online information ecosystem must account for user preferences and behavior, including their preferences for information for reasons other than its truth value (such as its usefulness for protecting worldviews and advancing interests) \cite{williams2022marketplace}. 
Very few users share articles from known fake news outlets \cite{grinberg2019fake,altay2022so}. 
One possible reason is that users perceive reputational harm from sharing information from unreliable sources \cite{altay2022so}. 
If the same belief can be supported with a mainstream source, rather than an unreliable source, it is likely to be preferred so as to avoid this reputational cost.
Articles by mainstream outlets that are co-shared with fake news (such as \textit{The Washington Post} article highlighted in Fig. \ref{fig:method_example} originally titled ``Vaccinated people now make up a majority of covid deaths'') might become a part of the rationalization process for misperceptions and misleading narratives or beliefs. 
Our case stud\replaced{ies}{y} suggest\deleted{s} as much, with tweets such as ``Well look what the CDC is saying today .... It's the vaccinated spreading covid.'' sharing the article \url{businessinsider.com/cdc-fully-vaccinated-people-can-spread-delta-variant-2021-7} with the headline ``CDC says fully vaccinated people spread the Delta variant and should wear masks: `This new science is worrisome'.'' 
The same misleading narrative can now be rationalized using a combination of different information sources (fake news, mainstream news, cross-partisan outlets). 

\replaced{Existing narratives present in misinformation content being also present}{The presence of existing misinformation narratives} in the mainstream media landscape also hints at the potential agenda-setting power of fake news. 
\added{Prior work has uncovered a temporal alignment in news volume and bursts of coverage between fake (or unreliable) and reliable news ecosystems, which suggests a competition for setting the agenda of public discourse \cite{budak2023bursts}.} 
Fake news content can both respond to, and influence, the agenda in more mainstream partisan media \cite{vargo2018agenda}, and mainstream news outlets are sometimes responsible for disseminating and popularizing particular fake news stories \cite{tsfati2020causes}. 
Another form of interaction between fake and mainstream media ecosystems is fake news outlets potentially capitalizing on the sensationalism \emph{already} present in the mainstream media environment \cite{guo2020fake}. 
As also highlighted by our study, this underscores the responsibility and need for structural action in the mainstream journalism landscape itself rather than confining the problem of misinformation to a certain set of identified `bad actors' in the form of specific fake news outlets. 

Our findings also complement recent work that more specifically examines false and misleading narratives regarding the COVID-19 vaccine within communities on social media \cite{sharma2022covid}. 
Not only are the identified narratives similar, but the authors in that work noted the ``distortion of facts due to misleading rather than outright false narratives'' such as exaggeration of side-effects and recontextualization. 
This mirrors our findings in terms of observed extracted narratives, the co-shared fake and mainstream news content, and the indication of strategic repurposing of mainstream news content on social media (which can help support efforts to recontextualize legitimate reporting). 
The latter indication of repurposing and possibly taking advantage of how headlines and mainstream story content are framed is backed by the findings of our case stud\replaced{ies}{y}. 
These findings also relate to recent scholarly understanding that the challenge of misinformation on social media is less about ``bad facts'' and more that people ``are more often misled not by false evidence but by misinterpretations and mischaracterizations --- dynamics of a collective sensemaking process gone awry'' \cite{starbird2023facts}.

\paragraph*{Limitations and Future Work.}

For our quantitative results in this study, we examine the content of articles, but for social media posts, we only considered the articles shared and not the text generated by the user while sharing the article. 
Our findings and our case stud\replaced{ies}{y} suggest that future work should also look at the \emph{content of tweets} sharing news articles, robustly identify misinformation sharers, and study \emph{how} mainstream news articles get repurposed to support pre-existing misperceptions via the potential choices made in tweet texts (while maintaining an understanding of the larger media ecosystem that enables such repurposing). 
Examining post content can also help verify what our findings indicate: mainstream outlets might be used to grant an air of legitimacy to certain narratives. 
Such an examination can help analyze the instances where mainstream news content itself misleads \emph{directly} and instances where mainstream news tends to \emph{enable} misleading usage by social media users. 
\added{Furthermore, while our approach in this work focuses only on external links shared on Twitter, false and misleading claims can and do often circulate in social media posts without external references or links; however, our approach is centered on the understanding that it is the credibility of the external reference that misinformation sharers capitalize on. }

Studying the content of the social media post sharing an article can also help assess if the article is being shared primarily to criticize the article, the outlet, or a particular type of media --- for example, by using ironical commentary such as `look at what the New York Times is saying now... .' 
To help assess the possibility of this `share-to-criticize' pattern in our dataset, we conducted a manual annotation analysis of a $100$ randomly selected tweets that share an article published by a trustworthy source, where the tweet is authored by a Twitter user with a history of sharing at least five distinct fake news articles. 
Three independent annotators labeled each tweet as to whether or not it was criticizing the article it was sharing (or if such a determination cannot be made without additional context). 
A majority of the annotators labeled the tweet as sharing the article to critique or dispute it (or the outlet or the type of media) in only 3 cases. 
This suggests that the `share-to-criticize' behavior is very limited in our dataset and unlikely to have influenced our results (more details on the human annotation task and analysis are provided in the \emph{SI Appendix}, section S10). 
However, future work should investigate the broader concern in which misinformation sharers share mainstream news articles on social media in a critical or ironic manner. 

Another subject of future analysis would be to examine where and how misleading narratives occur in mainstream news stories. 
One possibility is that these narratives do not occur in the content written by the journalist or news reporter themselves, but in \emph{direct quotations}: content attributed to a different actor or source replicated and used verbatim in a text. 
Such direct quotations are explicitly marked using quotation markers (such as ``...'') in texts \cite{pouliquen2007automatic}. 
We removed instances of direct quotations from co-shared and control articles in a preliminary analysis (using patterns that err conservatively toward removing rather than keeping potential directly quoted text), and found that our hypothesis still holds when working with this modified content. 
Therefore, the significantly higher presence of these potentially misleading narratives in co-shared articles does not seem to be \replaced{dependent on}{driven by} the practice of direct quotations in media alone. 
Overall, our effect holds for these newly processed texts (with direct quotations removed) in $16/18$ significance testing instances (where each testing instance corresponds to one row in \cref{tab:misinfo_results_sig_testing}). 
We report our preliminary analysis including the procedure to remove direct quotations and detailed results in \emph{SI Appendix}, section S5.

Nonetheless, future work should examine whether the practice of direct quotations (and the possible presence of misleading narratives in those quotes) necessarily abdicates the responsibility of journalists. 
Direct quotations still constitute a journalistic choice; indeed, one prior work argues that journalistic quotation (direct or indirect), is a `highly interpretive compositional activity', making it a subjective part of creating a news story \cite{harry2014journalistic}. 
Direct quotations can constitute a \emph{shift} in responsibility to the source as well as a shift in the interpretation or attitude towards the quoted content as not a part of the factual reporting of events \cite{bergler2006conveying}. However, even if this is the case, it is possible that in the context of strategic repurposing of mainstream news, misinformation conveyors, as well as their audience, might be able to ignore this shift. Future work can also examine the effect of indirect quotation, where quotation markers are not used and instead, quotes are presented using reporting verbs such as `said' (`they \emph{said} that $\dots$ \emph{reported content} $\dots$'). 
Such an investigation can help further understand the distribution of where misleading narratives occur in mainstream media reporting --- when citing a source verbatim or in original writing by the author of the news article. 

Our work uses one particular way to categorize outlets as fake news --- outlets annotated as `regularly publishing false content' --- aiming for precision. Future work can experiment with a broader set of unreliable content to study the interactions between reliable and unreliable content.


By scoring individual news articles for their co-sharing likelihood, our work can also help inspire alternatives to the dominant research paradigm that uses domain-level reliability ratings and tends to treat all individual articles published by the same outlet in the same manner. 
We note that we also treat articles published by \emph{fake} news outlets in that manner, but we consider each individual story published by \emph{mainstream} outlets to identify cases of misinformation adjacency. 
We also encourage future work to further refine the set of potentially misleading narratives obtained and used in this study to help answer specific research questions and conduct additional investigations of specific misleading narratives. 
In Tables \replaced{S46 to S66}{S15 to S35}, we provide all the high-dimensional narratives automatically extracted from the texts of fact-checked false claims and used in the comparison of their presence in co-shared and control mainstream articles (for the results reported in \cref{tab:misinfo_results_sig_testing}).

While we quantify the relative presence of potentially misleading narratives in the co-shared and control groups of mainstream articles, our focus in this work is to compare and test the \emph{difference} among the two groups, and not quantify the presence of possible misinformation in various mainstream media articles. The specific values for the presence of potentially misleading narratives in mainstream articles rely on the method itself, which in turn relies on keywords and extracts many narrative structures from any piece of text. To quantify the presence of potentially misleading narratives in mainstream media more accurately, future work would need to propose a more precise method for that particular quantitative estimation, which is out-of-scope for our work. That future work can certainly build on our work, especially our research design, to identify articles of interest in the quest for quantifying the amount of misleading framing in misinformation articles.

\added{The narrative extraction tool we use first splits the corpus into sentences, and then works at the sentence-level. This limits the context to just the sentence, which can result in some artifacts. For example, while negations within a sentence are captured, if someone presents a claim in an article only to then negate it later on in a different sentence, this will not be captured by the method.}

Finally, we note that contending with other explanations can clarify the relative role of potentially misleading narratives and also serve as robustness checks for our work. 
Here, we accounted for the \replaced{potential relevance}{effect} of partisan curation of news articles (\emph{SI Appendix}, section S2). 
However, other possible distinctions between the co-shared set of articles and our control group can be the relative presence of certain entities, or even interest in specific topics that may dominate the fake news world and co-shared mainstream articles. 
Conducting our hypothesis tests within subsets of articles based on the topic or theme in the news content, and also subsets defined by the presence of prominent entities, constitute interesting directions for future research enabled by our processed dataset. 

\section*{Methods}

The Twitter dataset used in this study was previously approved for research \cite{hughes2021using,shugars2021pandemics} and collection by the Institutional Review Board (IRB) at Northeastern University (17-12-13). 

\paragraph*{Original Twitter data: Construction of the underlying Twitter dataset and data collection.} 

Our initial dataset consists of tweets collected for a panel of over $1.6$ million U.S. Twitter users that are matched with their voter registration records, obtained via a commercial voter file provided by TargetSmart (described in prior work \cite{hughes2021using}). 
This panel slightly over-represents white and female Twitter users, but is otherwise demographically similar to samples of Twitter users collected via high-quality surveys \cite{hughes2021using,shugars2021pandemics}. 
This panel is appropriate for our study as we are chiefly concerned with (mis)information shared on social media, for which the sharing behavior of users is necessary and the ability to verify that these users are real people is extremely useful. 
The sample was constructed by identifying users active in the Twitter Decahose data (a random $10\%$ sample of all of Twitter that was available to academics via the Twitter API before 2022) with a unique first name/last name/location that also appeared in a commercial voter file. In other words, Twitter users were exact-matched to a commercial voter file by first name, last name, and location. 
All matches were retained in the panel (\emph{i.e.} we did not randomly sample). After the user panel was constructed, users’ tweets were collected at regular intervals using the Twitter API. 
Further details about construction and validation of this dataset are provided in prior work \cite{hughes2021using,shugars2021pandemics}. 

We retain English tweets that share news articles (published by English-language information sources or outlets per \newsguard{}) from May 1 2018 to November 14 2021. 
This results in an initial dataset containing ${\sim}14.6$ million unique \abr{url}s published by ${\sim}4000$ domains shared by ${\sim}600,000$ different users. 
For each tweet, we have the user \abr{id} of the author, and the news \abr{url} being shared in the tweet. 
We retain \abr{url}s that were shared by at least $20$ different users (to control for sparsity effects; we did not conduct further sampling as our study is large-scale observational), resulting in ${\sim}420,000$ different \abr{url}s, published by ${\sim}2400$ domains, and shared by ${\sim}450,000$ different users. 
Compared to the full panel of about $1.6$ million U.S. Twitter users, our subset of ${\sim}450,000$ users slightly overrepresents white and female Twitter users, and overrepresents users registered as Democrats. Our subsample also slightly underrepresents Twitter users in the $18-29$ age bucket and, in almost equal proportion, slightly overrepresents users in the $50-64$ age bucket. 

Our choice of date range was influenced by two internal criteria: 

\begin{enumerate}
    \item Use the latest Twitter panel data at the time of the beginning of experimentation and research design (which began in mid 2021, and was revisited again at the end of 2021 when our methodology took concrete shape). 

    \item Compute-related restraints, such as memory, for loading and working with large datasets. 
\end{enumerate}

This is how our date range ended up being May 2018 to November 2021: we tried to go back in time from November 2021 as much as computationally feasible. The reasoning behind trying to work with the latest data at the time of initial experimentation was to retain the latest narratives and shared news in a fast evolving political climate. Our date range encompasses the whole of 2020, which was a unique year due to the onset of a once-in-a-century pandemic in COVID-19, and also contained the U.S. presidential election, seeing an active and charged political climate throughout. As such, it is unsurprising that 2020 news stories seem to feature prominently in our work. 

As noted above, the data for this study comes from a pre-existing panel of Twitter users matched to a commercial voter file. The commercial U.S. voter file was obtained from TargetSmart in October 2017. Twitter API v1 was used to collect Twitter data. The text data of news articles was collected using the Newspaper 3k API (version 0.2.8) in Python as a Python library (\url{newspaper.readthedocs.io/en/latest/}). \newsguard{} data (another commercial resource) provided classifications of domains as detailed below.

\paragraph*{Use of \newsguard{} annotations.}

\newsguard{} is an independent organization that involves a team of journalists evaluating the quality of online information sources. \newsguard{} uses web tracking data along with ratings by professional news editors and journalists (a trained team of experts) to produce the final sets of annotations and scores at the domain level. The panel of editors and journalists offer a trustworthiness score between $0$ and $100$ which is based on several different criteria, and annotations for those criteria are also included. These criteria have to do with journalistic quality and standards of transparency and credibility. As such, the criteria include publishing of false information, fact-checking standards of the news institution, disclosing advertising agencies and all sources of funding, being transparent about ownership, level of editorial oversight, etc. Having annotations for these different criteria and not just a score or a `blocklist' of domains is useful in making finer-grained assessments and allows for nuance. For example, less-reliable domains can be distinguished on the basis of the potential reason for a lack in quality: not disclosing ownership details is qualitatively a different reason to not trust a website than the regular publishing of false information. Detailed descriptions of all journalistic criteria involved in the ratings are available at \url{web.archive.org/web/20240131155450/https://www.newsguardtech.com/ratings/rating-process-criteria/}. 

\newsguard{}'s domain-level judgments show a high overlap with other lists with publisher ratings \cite{allcott2019trends}; specifically, recent work comparing six sets of news domains' quality ratings (including \newsguard{} ratings) found that ``they generally correlated highly with one another'' \cite{lin_lasser_lewandowsky_cole_gully_rand_pennycook_2022} (see also \cite{guess2021cracking,edelson2021understanding}). 
\newsguard{}'s ratings and annotations are commonly used in computational misinformation research \cite{guess2021cracking,edelson2021understanding,luhringbest}, albeit in different ways --- such as using the provided `trustworthiness' scores to divide information sources into groups \cite{guess2021cracking,edelson2021understanding}. \newsguard{} has sought to add news sources over time, and its popularity in misinformation research warrants an independent investigation of its content, processes, and use. This is the goal of a recent article \cite{luhringbest}, with the manuscript made publicly available as a resource examining \newsguard{} annotation data in detail: \url{osf.io/preprints/psyarxiv/v6e4b}.

In this work, we aim for precision by considering `regularly publishing false content' as our criterion for demarcating fake news outlets, out of all the factors that determine trustworthiness in \newsguard{} ratings. This helps highlight the drawbacks of conceptualizing problems with the digital information ecosystem as that of ‘false information’ or fake news content — a framework that often foregoes closer examination of information coming from sources deemed reliable. All other outlets are deemed `reliable'. We used this specific annotation for domains in \newsguard{} data to obtain fake news domains. Since one common way prior work has used \newsguard{} data is the provided `trustworthiness' scores for outlets \cite{guess2021cracking,edelson2021understanding,luhringbest}, we also ran our experiments for trustworthy domains, assessing the effects of a different classification, and showed that our findings are robust to this particular way of filtering for reliable outlets (Table \ref{tab:misinfo_results_sig_testing}). 

We rely on \newsguard{} annotations because of their particular fit with our research questions, since to the best of our knowledge, the other annotated datasets or manual annotations of news domains do not provide annotations for `regularly publishes false information'. 
There is also a broader issue with different domain annotation efforts annotating for different definitions and slightly different phenomena; tackling this challenge is out-of-scope for our work. This challenge along with the issue of list obsolescence complements our broader concerns regarding domain-level categorizations and the need to study articles and individual stories from all domains on their own terms. Our work takes a step in that direction, but as one of the first works to do so, is still somewhat shackled by the precedents set in the field including use of particular resources such as \newsguard{}. We support parallel efforts to create publicly available lists for domain and story quality, and our work provides ideas on how to take advantage of existing annotations to move beyond the domain-level and identify adjacent news stories from mainstream domains that do not often receive attention in misinformation research. Our method and our codebase can help other researchers to take advantage of social media based co-sharing, while our qualitative case studies can offer journalists and scientists insights into how `technically factual' information nevertheless plays a role in the broader misinformation landscape.

\paragraph*{Obtaining set of \emph{mainstream} outlets.}

We borrow domain-level ratings from prior work~\cite{greencuration}. 
These ratings are based on Twitter shares of articles (over the $2010$ to $2021$ period) published by various outlets using the same panel of users as in the overall dataset used in this work (these particular domain-level annotations are made publicly available as part of our codebase). 
The two criteria we use to demarcate mainstream outlets are detailed below: 

\begin{enumerate}

    \item \textbf{Publish predominantly political content.} Each domain publishes many articles, but only some of them were deemed as `political' in domain-level ratings based on the type of users sharing them on Twitter. Using the percentages of all articles published by a domain deemed political content as the political score, \emph{predominantly political domains} are ones having a score greater than a $z$-score of $1.0$ (greater than the mean plus the standard deviation calculated using all the domain-level political scores). 
    
    \item \textbf{Popular.} Among these domains that predominantly publish political content, we select the \emph{popular} ones as those that are \emph{i)} in the top $10\%$ for the number of Twitter users sharing a domain's content; and \emph{ii)} among the top $500$ most-visited domains per the users involved in a YouGov panel survey (2015-16), with site-visiting statistics derived from passive metering data (per the analysis and data-release in prior work \cite{guess2021almost}). Using this intersection is a more robust way of deciding the domains that should be deemed `popular' as opposed to, say, only considering popularity based on Twitter shares. 

\end{enumerate}

\paragraph*{Audience-based partisanship scores for news outlets.}
Our Twitter-based dataset contains a model-based likelihood of identifying as a Democrat at the user level. 
These scores are used to estimate the average partisan lean of each \abr{url}.
The scores are normalized to lie within a $[-1, 1]$ (Democrat-Republican) range (in the vein of the traditional \emph{negative-to-positive} as \emph{liberal-to-conservative} axis). 
These Twitter audience-sharing-based scores are aggregated to obtain outlet or domain-level scores following prior work \cite{greencuration}. 
We use the median of these partisanship scores to classify the outlets as liberal or conservative.

\paragraph*{Creating two groups of mainstream news articles: disproportionately co-shared and control.}  

Our process for finding co-shared mainstream articles is visualized in Fig. \ref{fig:method_example}b. 
We construct a weighted graph with individual articles or \abr{url}s as nodes. 
Edges between a fake news \abr{url} (nodes shown on the left-hand side of Fig. \ref{fig:method_example}b) and a reliable news \abr{url} (nodes shown on the right-hand side of Fig. \ref{fig:method_example}b) are weighted by the number of Twitter users in our dataset that shared \emph{both} the incident node \abr{url}s at least once. 
For example, in Fig. \ref{fig:method_example}b, $8$ is the edge weight since eight different users in our panel shared both \textit{The Washington Post} article and the fake news article displayed. 
We only consider edges between a pair of nodes where one node is a fake news article and the other node is a reliable article. 
Our processed dataset contains ${\sim}24k$ fake news stories published by $276$ domains.

We apply a graph-pruning algorithm \cite{dianati2016unwinding} to assign each edge a co-sharing score based on the likelihood of both incident node \abr{url}s getting shared by the same user, controlling for the individual popularity of each of those \abr{url}s in isolation. 
Specifically, we use the expected likelihood of two articles getting shared by the same group of people, based on how many people share each of those articles independently, as a null model for determining the actual co-sharing likelihood. 
For example, an edge assigned a weight of $10$, representing $10$ people sharing both article 1 and article 2, will receive a higher co-sharing score if article 1 and article 2 each get independently shared by about $20 - 30$ people, compared to them getting shared by $200 - 300$ people each. 

Graph-pruning methods have previously been applied to study the media landscape by constructing a domain-level co-exposure network that identifies news outlets with a shared audience \cite{grinberg2019fake,majo2019backbone}. 
However, by focusing on a network of \emph{individual} news shares, we can investigate mainstream news stories published by reliable outlets that would otherwise not be considered for their potential role in misinformation circulation on social media. 
In our implementation of a graph-pruning algorithm on a graph of shared URLs, edges in the top $1\%$ of co-share scores are considered disproportionately co-shared or simply, co-shared (highlighted as the darker edges in Fig. \ref{fig:method_example}b). Following recommendations and practices in prior work \cite{dianati2016unwinding,grinberg2019fake}, we set a ``high threshold for deviation from the null model.'' 
To obtain the 0.99 quantile (or top $1\%$) value to set the threshold, we draw $1000$ random samples of $1000 \times 1000$ sized subgraphs from the entire graph and use the distribution across the samples to estimate the threshold value with $95\%$ confidence. 
The sampling procedure is used for estimation since using the entire $\sim 24,000 \times \sim 396,000$ graph proved to be computationally intractable (where $\sim 24,000$ represents number of fake news articles, and $\sim 396,000$ represents number of articles from reliable outlets). 

We consider articles published by mainstream outlets in this top $1\%$ set of articles as \emph{the group of mainstream news articles that are co-shared with fake news}. 
Fig. \ref{fig:top_nyt_wapo_coshared} displays examples of top co-shared articles published by The \textit{The New York Times} and the \textit{The Washington Post}, and similarly Fig. S1 in \emph{SI Appendix} displays top co-shared articles published by CNN and Fox News. 
We note that mainstream outlets form more than $90\%$ of all reliable outlets resulting in about $2000$ domains publishing a total of ${\sim}395k$ \abr{url}s in our dataset.  
Stories published by the same set of mainstream outlets that are \emph{not} in the top $5\%$ of co-sharing likelihood scores form our \emph{control group}. 

Finally, we have ${\sim}91k$ news stories in the co-shared group and ${\sim}51k$ news stories from the same set of news outlets in our ``control" group --- i.e. mainstream articles that are not connected with any fake news articles in the top 5\% of all likelihood scores. 
This is a strict control criterion, separating the co-shared and control sets of articles, and explains the lower number of control articles (as opposed to what would be expected when simply considering every other article not in the co-shared set as the comparison group). 
Apart from these threshold settings of top $1\%$ for co-shared group, and never in top $5\%$ for control group, we also run eighteen different experiments with different threshold settings,  across top $2\%$, top $1\%$, top $0.5\%$, and top $0.1\%$ (for co-shared articles) --- compared with below $5\%$, below $4\%$, below $3\%$, below $2\%$, and below $1\%$ (for control articles). This is discussed in Section S8 in the \emph{SI Appendix}, with the replicated results across different thresholds presented in Tables S27 to S44. We find that our main results continue to hold for narratives extracted from the content of fake news articles, replicating across the board. There are partial differences for some settings such as using top $3\% - 1\%$ for the control set, which informs researchers of the sensitivity of our results to some threshold settings: for absolutely no change to our main results,  one should use top $2\%$ or top $1\%$ of co-sharing scores for assigning articles to the co-shared group, and articles never in the top $5\%$ or top $4\%$ co-sharing scores for assigning articles to the control group.

News article content (headline and story text) of all \abr{url}s across the various groups of articles --- fake news, co-shared mainstream news, and the control group --- are automatically extracted. 
We scrape contents of news articles using the \abr{url}s to get the headline, subheading, and story texts along with the publication date using the \textsc{Newspaper3k} \abr{api}, available as a Python library (\url{https://newspaper.readthedocs.io/en/latest/}). 
Our code as well as our entire collection of texts are provided as part of our codebase, though the Twitter data itself will be shared in a restricted manner following Twitter and \abr{irb} guidelines.

\paragraph*{Narrative extraction.}

We operationalize narratives using methods and an open-source package (\emph{relatio}) developed in prior work \cite{ash2023relatio}, available at \url{github.com/relatio-nlp/relatio}; we used version 0.2.1 of \emph{relatio} package in Python for this study.  
The conceptual definition used is ``sets of relationships between entities that act on each other'', which, in language, involves looking at ``grammatical statements describing actors, actions, and the acted-upon.'' 
Fig. \ref{fig:narr_extraction} briefly visualizes the process of extracting the narrative structure through an illustrative example. 

Narrative structures extracted consist of the base form: $$\text{AGENT ENTITY} \xrightarrow{\text{VERB}} \text{PATIENT ENTITY} \in E \times V \times E$$ where $E$ is the set of clustered entities and named entities, and $V$ is the set of all verbs. 
For example, consider a sentence present in one article in our set of fake news: ``Zostavax vaccine caused multiple people to develop shingles.'' 
In this case, $\mathit{vaccine} \xrightarrow{\mathit{cause}} \mathit{shingles}$ and $\mathit{people} \xrightarrow{\mathit{develop}} \mathit{shingles}$ could be extracted as these base structures, and we can also have $\mathit{vaccine} \xrightarrow{\mathit{cause}} \mathit{people} \xrightarrow{\mathit{develop}} \mathit{shingles}$ (extracted narrative structures need not be limited to triplets). 
These are then simply represented as strings, or what we call narrative labels or narrative strings (`vaccine cause shingles'; `vaccine cause people develop shingles'). 

We note that the automated output of this package can be noisy, and not all extracted strings themselves will correspond to a meaningful notion of `who did what to whom'. 
For this reason, the original paper referred to automatically extracted statements as narrative \emph{candidates} \cite{ash2023relatio}. 
We refer to these outputs as narrative strings or narrative labels, and sometimes simply as \emph{narratives} in this paper, but for fake news articles containing many sentences per article, our work considers \emph{recurring} narratives as potentially misleading narratives.

As shown in Fig. \ref{fig:narr_extraction}, entities are clustered so that variants of the same entity receive the same entity label. 
Clustering represents the dimensionality reduction step for narrative extraction, since large collections of texts will refer to the same entity in a variety of ways, which includes grammatical variation (`vaccine' and `vaccines') as well as various ways to refer to the same named entity (`trump' and `donald trump'). 
The pre-specified number of clusters for clustering entities during narrative extraction (Fig. \ref{fig:narr_extraction}) constitutes an important decision. 
Two settings are used to extract narratives at two levels of granularity from the text data (following \cite{ash2023relatio}): \emph{low-dimensional narratives}, allowing fewer, large-sized clusters or latent entities that result in `broader' narratives, and \emph{high-dimensional narratives}, allowing a higher number of smaller-sized clusters that result in more specific or `fine-grained' narratives. 
For example, for the (original, non-processed) sentence --- ``The filing claims that Zostavax caused multiple people to develop shingles.'' --- the method finds the low-dimensional narrative `vaccine cause shingle', and finds the high-dimensional narrative `zostavax cause multiple people develop shingle'. 
Each instance of our hypothesis testing and corresponding results includes both these levels of granularity, adding another layer of robustness to our findings. 
We note that verbs, identified as actions connecting entities in the semantic role labeling step, are not clustered in the narrative extraction method as opposed to entities (subject and patient). 
This is primarily due to the difficulty of reliable embedding-based clustering for verbs, where verbs implying opposite actions can get clustered together \cite{ash2023relatio}. 
This has to be avoided since it could lead to opposing narratives (such as `vaccine harm children' and `vaccine helps children') being folded under the same narrative. 
Therefore, all verbs are treated independently and no dimensionality reduction is performed on them \cite{ash2023relatio}. 


The paper introducing the narrative extraction tool, called \emph{relatio} and open-sourced as a Python package, also provides a visualization of the complete algorithm underlying the method in their Figure 2 \cite{ash2023relatio}. Here we describe the algorithmic process underlying the method in brief that goes from the corpus to extracted narrative statements (these steps are not always sequential and can occur in parallel). 

\begin{itemize}
    \item Step 1: The corpus is split into sentences, and the entire narrative extraction process occurs at the sentence-level. 

    \item Step 2: Named entity recognition (NER) is run on the sentences to identify named entities. 

    \item Step 3: Semantic Role Labeling is used to identify roles for various entities, including agents, patients, and attributes. Semantic Role Labeling also helps identify instances of verb being negated (in order to retain, for example, `not winning' as distinct from `winning' as a verb being negated flips the narrative). 

    \item Step 4: The combination of steps 2 and 3 provides semantic roles with named entities, and roles without named entities. 

    \item Step 5: Sentences (output of step 1) are used to train and fine-tune a embedding model (that can work on words and phrases). 

    \item Step 6: The embedding model (step 5) is also trained and fine-tuned with the extracted semantic roles without named entities (step 4) to make role embeddings. 

    \item Step 7: The role embeddings (step 6) are used to train K-means clustering algorithm and label clusters. 

    \item Step 8: Finally, the extracted roles with named entities (step 4), the verb negations (step 3), and the label clusters (step 7), together result in narrative statements or narrative labels or narratives. 
\end{itemize}

Narratives extracted using this tool from speeches on the floor of U.S. Congress (both the House and the Senate, 1994-2015) have been shown to `reflect key events in U.S. history', to provide a `qualitative window' into the priorities and values of U.S. congress members, and to help map ideological disagreements \cite{ash2023relatio}. Additionally, the network of entities and the links between them (in form of verbs or actions) exposed the connections between various topics of debate on the floor of U.S. Congress. While floor speeches and news articles are different domains of text, the comprehensive range of analyses enabled by this tool in one setting where texts tend to follow some structural similarity (as is the case for floor speeches and for news articles), and crucially, the application of tool in diverse contexts such as social media posts \cite{ash2023relatio,sipka2022comparing}, as well as our work's domain of news coverage \cite{ottonello2022financial}, all together provides assurance for its use as a component in our approach and analysis.

\paragraph*{Creating libraries of potentially misleading narratives.}
In this work, we combine all the texts involved in our method: fake news articles, false fact-checked claims, and co-shared and control sets of mainstream news articles, and apply the narrative extraction tool to this combined text data. 
This ensures that various choices in narrative labels or string representations, such as the chosen entity representation for a cluster, remain consistent across the dataset. 
We then divide the output according to the different types of data being considered to get the set of narratives for each group of texts. 

For potentially misleading narratives, we consider two sources of data. 
The first consists of fake news articles (${\sim}24k$ \abr{url}s), where we consider the headline and story for each article as separate text instances, since many (at least 6 out of 10) people on social media only read the headline of an article \cite{gabielkov2016social}; see also: \url{americanpressinstitute.org/publications/reports/survey-research/how-americans-get-news/} and \url{washingtonpost.com/news/the-fix/wp/2014/03/19/americans-read-headlines-and-not-much-else/}. 
 Headlines and story texts are deduplicated and processed to remove information like domain name mentions and headline text within the story text. 
This results in ${\sim}30k$ unique texts derived from fake news articles. 
The second source consists of texts of claims that have been fact-checked as false (${\sim}24k$ unique texts or claims). 
Potentially misleading narratives from fake news content are defined as recurring narratives: the top $1\%$ in terms of frequency of occurrence (therefore, these narratives occur across several texts in our dataset). We also experiment with all the narratives found in fake news content (Table \ref{tab:misinfo_narr_libs}). 

For fact-checked false claims, we collect and combine various publicly available datasets. 
Specifically, sources of false fact-checked claims include \emph{a)} claims fact-checked by PolitiFact, The Washington Post, and FactCheck.org as collected by the Data Commons initiative (\url{datacommons.org/factcheck/download\#research-data}); \emph{b)} fact-checked public health claims \cite{kotonya2020explainable}; \emph{c)} a benchmark dataset of fact-checked claims from PolitiFact \cite{wang2017liar}; and \emph{d)} a large-scale dataset of fact-checked claims coming from $26$ different fact-checking websites \cite{augenstein2019multifc}. 
All claims are naturally occurring and in the English language. 
These publicly available claim texts alongside fact-check labels are combined; this combined claims data, along with the process of filtering to retain only false claims, is documented and provided in our codebase. 

We use the claims that are labeled as false (or an equivalent label such as `not true'; our codebase provides all the labels used to create our collection). 
These claims are provided along with the rest of our data in our publicly available codebase. 
Since these are statements, usually consisting of a single sentence fact-checked as false, we consider \emph{all} narratives extracted from this data as potentially misleading narratives. 

The above process ultimately results in three potentially misleading narrative libraries, each containing low- and high-dimensional narratives (Table \ref{tab:misinfo_narr_libs}). 
Table \ref{tab:misinfo_narr_examples} presents examples of potentially misleading narratives.

\paragraph*{\added{Hypothesis testing setup.}}

Consider a particular library of potentially misleading narratives and a particular set of high- or low-dimensional narrative labels or strings in that library. Let us denote this particular collection as $M$, with size $|M|$ and narrative strings $n_1, \dots, n_i, \dots, n_{|M|}$. 
For both groups of mainstream articles (co-shared and control), we obtain the corresponding values of each potentially misleading narrative as its normalized count in the set of all narratives extracted from that particular group of articles. 
Specifically, let $C_s$ be the collection of all narrative strings present in co-shared mainstream articles (with $|C_s|$ total number of unique narrative strings: $s_1, \dots, s_i, \dots, s_{|C_s|}$), and $C_o$ be the collection of all narrative strings present in control mainstream articles (with $|C_o|$ total number of unique strings: $o_1, \dots, o_i, \dots, o_{|C_o|}$). 
Then, the two lists of values we compare --- $l_{cs}$ for co-shared, and $l_{co}$ for control --- are created by iterating over all narratives ($n_i$) in $M$:
$$\displaylines{ 
l_{cs} = \left[\frac{\text{count}(n_i \in C_s)}{\sum_{i = 1}^{|C_s|} \text{count}(s_i \in C_s)}  \forall n_i \in M \right]\cr 
l_{co} = \left[\frac{\text{count}(n_i \in C_o)}{\sum_{i = 1}^{|C_o|} \text{count}(o_i \in C_o)}  \forall n_i \in M \right]
}
$$
We then test if potentially misleading narratives have a significantly higher presence in co-shared articles ($l_{cs} > l_{co}$; Fig. \ref{fig:method_example}c) using the one-sided Wilcoxon-signed rank test. 
The same potentially misleading narrative labels ($M$) are measured for their counts or presence across articles in both groups.

\paragraph*{\added{Note on randomization.}} 
The only part of our core methodology as developed in this study which involves randomization is the estimation of quantile values (the 0.999, 0.995, 0.99, 0.98, 0.97, 0.96, 0.95 quantile values) during our graph-pruning based method of computing co-sharing scores between news from reliable outlets and fake news outlets, in order to set thresholds for the co-shared and control groups. As specified above, we draw $1000$ \added{random} samples of $1000 \times 1000$ sized subgraphs from the entire graph and use the distribution across the samples to estimate the threshold value with $95\%$ confidence. Random sampling in our study is only driven in this particular component by computational feasibility concerns. This random sampling is implemented via \texttt{random} module or library in Python, which uses the Mersenne Twister as the core generator of pseudo-random numbers \cite{matsumoto1998mersenne}. 

Randomization is not being used for any group allocation, and therefore, the concept of blinding is not relevant to this study.

\subsection*{Data Availability}

We use publicly available tweets and the text of published news articles. 
While our public, no-restriction data sharing ability is limited by the restrictions placed on data sharing by Twitter’s Terms of Service, broader privacy concerns regarding the panel of users, and data-use agreements for NewsGuard and TargetSmart data, we make aggregated data files along with all the code required for replicating our findings freely and publicly available at \url{zenodo.org/records/15182977} (DOI: \url{doi.org/10.5281/zenodo.15182977}). These data files include all derived data used in our research for testing our hypotheses and everything necessary for replication of main results (Table \ref{tab:misinfo_results_sig_testing}). 
The only parts of data not made available in this public, no-restriction setting are: 

\begin{enumerate}
    \item Raw data about the tweets (i.e. user IDs, tweet IDs, tweet texts, etc.), restricted following Twitter policy. While sharing of Twitter data is complicated by Twitter's evolving policy on researcher access, per our interpretation of latest terms of service from Twitter, tweet IDs can also not be made publicly available. The lack of public availability is also due to user identification concerns for our Twitter panel. 

    \item NewsGuard-based data that could directly reidentify their annotations (such as groupings of domains, which will directly reveal NewsGuard annotations), in line with legal agreement with NewsGuard (a proprietary data source). 

    \item U.S. voter file data acquired from Targetsmart in 2017 (with which the Twitter panel was constructed and which ensures the presence of all real users in our data, necessary to investigate our research questions). This proprietary data cannot be directly shared in raw form per legal agreement with Targetsmart. 
\end{enumerate}

However, Twitter data as well as the data \emph{derived} from the two proprietary resources used in this research are also made available to researchers who apply for access and sign a data use agreement, following the precedent set in prior work \cite{grinberg2019fake}. This agreement will explicate the requirement that no attempts be made ``to identify, reidentify, or otherwise deanonymize the dataset or the restricted resources'', and restrict further sharing until explicit approval is obtained from Northeastern University. This is necessary both from a user privacy point of view as per Northeastern’s IRB approval for the dataset (17-12-13), as well as legal agreements made with TargetSmart and NewsGuard. Researchers can contact the corresponding author of this study or David Lazer at Northeastern University to obtain access to restricted data. 

For fact-checked false claims, we collect and combine various publicly available datasets. 
Specifically, sources of false fact-checked claims include \emph{a)} claims fact-checked by PolitiFact, The Washington Post, and FactCheck.org as collected by the Data Commons initiative (\url{datacommons.org/factcheck/download\#research-data}); \emph{b)} fact-checked public health claims \cite{kotonya2020explainable}; \emph{c)} a benchmark dataset of fact-checked claims from PolitiFact \cite{wang2017liar}; and \emph{d)} a large-scale dataset of fact-checked claims coming from $26$ different fact-checking websites \cite{augenstein2019multifc}. 
All claims are naturally occurring and in the English language. 
These publicly available claim texts alongside fact-check labels are combined; this combined claims data, along with the process of filtering to retain only false claims, is documented and provided in our publicly available data and codebase. 



\subsection*{Code Availability}


All of our code, along with the aggregated dataset files needed to replicate our main findings in Table \ref{tab:misinfo_results_sig_testing}, is available at \url{zenodo.org/records/15182977} --- under the Creative Commons Attribution 4.0 International license. 
Custom code using Python and R was used in all analyses. 

\section*{Acknowledgments}
We thank Alexander Hoyle, Peter Rankel, Yi Ting, Stefan McCabe, Briony Swire-Thompson, Sarah Shugars, Kenny Joseph, Nir Grinberg, and Alan Mislove for their helpful technical support and feedback. We thank Alyssa Smith, Ata Uslu, Alexi Quintana Mathé, Allison Wan, Burak Özturan, Sagar Kumar, and Yukun Yang for support with manual annotation and constructive discussions. This research was enabled by material resources provided by \newsguard{} and TargetSmart.

\subsection*{\deleted{Funding:}}

\added{Funding: }

National Science Foundation award 2008761 (PG, PSR)
\\
National Science Foundation award 2031736 (PG, PSR)
\\
Hewlett Foundation grant \#2019-9224 (DL)
\\
Volkswagen Foundation grant ``Reclaiming Individual Autonomy and Democratic Discourse Online'' (DL)

\added{The funders had no role in study design, data collection and analysis, decision to publish or preparation of the manuscript.}

\section*{Author contributions statement}

Conceptualization: PG, JG, DL, PSR
\\
Methodology: PG
\\
Software: PG
\\
Formal Analysis: PG
\\
Investigation: PG, JG
\\
Visualization: PG, PSR
\\
Supervision: JG, DL, PSR
\\
Writing—original draft: PG
\\
Writing—review \& editing: PG, JG, DL, PSR
\\
Funding Acquisition: DL, PSR

\section*{Competing interests statement}

\replaced{The authors declare no competing interests.}{Authors declare that they have no competing interests.}

\clearpage

\section*{Tables}

\begin{table*}[!b]
\centering
\caption{\textbf{Examples of extracted narratives.} These are the most frequent high-dimensional narrative strings or labels (and sentences containing those labels) in misinformation content, that are present in co-shared articles but not present in control articles published by mainstream `real' news outlets. Terms presented in boldface in sentence text highlight the narrative label's presence in the sentence.}
\resizebox{0.96\textwidth}{!}{%
\begin{tabular}{@{}ll@{}}
\toprule
\multicolumn{1}{c}{\textbf{Narrative label}} &
  \multicolumn{1}{c}{\textbf{Sentences containing narrative label}} \\ \midrule
baby survive abortion &
  1. the data on \textbf{babies surviving abortions} is compiled  on a mandatory basis by only a handful of states. \\[0.15cm]
 &
  \begin{tabular}[c]{@{}l@{}}2. should a \textbf{baby survive} an \textbf{abortion}, the health care practitioner shall ``exercise the same degree of professional skill, \\care, and diligence to preserve the life and health of the child as a reasonably diligent and conscientious health care \\practitioner would render to any other child born alive at the same gestational age.''\end{tabular} \\[0.33cm]
 &
  \begin{tabular}[c]{@{}l@{}}3. 77\% of americans want to protect \textbf{babies} who \textbf{survive abortions}, nancy pelosi has blocked a vote 80 times\end{tabular} \\[0.15cm]
 &
  \begin{tabular}[c]{@{}l@{}}4. but data from minnesota, the centers for disease control and other sources indicate that \textbf{babies} do \textbf{survive abortions}, \\and laws are needed to protect them.\end{tabular} \\[0.15cm]
 &
  \begin{tabular}[c]{@{}l@{}}5. not only are women given the right to sue abortionists for the death of their \textbf{babies} that \textbf{survive} \textbf{abortions}, those \\in violation can be criminally charged in federal court for intentionally killing or attempting to kill a human being.\end{tabular} \\ \hline
russian hack dnc &
  1. stone claims if the \textbf{russians} didn't \textbf{hack} the \textbf{dnc}, his case should be dismissed. \\[0.15cm]
 &
  \begin{tabular}[c]{@{}l@{}}2. the mueller report clearly states that \textbf{russians hacked} the \textbf{dnc} and gave the hacked emails to wikileaks:\end{tabular} \\[0.15cm]
 &
  \begin{tabular}[c]{@{}l@{}}3. trending: the gateway pundit and american gulag donate \$40,000 to persecuted jan. 6 families this christmas on \\march 8, 2020 and before on june 16, 2019, we presented arguments against the mueller gang's assertion that \\the \textbf{dnc} was \textbf{hacked} by \textbf{russians}.\end{tabular} \\[0.33cm]
 &
  \begin{tabular}[c]{@{}l@{}}4. the news magazine claimed crowdstrike was correct in assessing that the \textbf{dnc} was \textbf{hacked} by \textbf{russians} in 2016.\end{tabular} \\[0.15cm]
 &
  \begin{tabular}[c]{@{}l@{}}5. when this news hit the media like a bombshell out-of-control left wing judge amy berman jackson tightened the \\unconstitutional gag order on roger stone to prevent him from discussing the shocking revelation and kravis filed a \\sur reply with the court falsely claiming that the us government and the mueller investigation had additional \\evidence to bolster their claim that the \textbf{russians hacked} the \textbf{dnc}.\end{tabular} \\ \hline
bill stop infanticide &
  1. this is the second time senate democrats have blocked the \textbf{bill} to \textbf{stop infanticide}. \\[0.1cm]
 &
  2. house democrats again block request to vote on \textbf{bill} to \textbf{stop infanticide} \\[0.1cm]
 &
  \begin{tabular}[c]{@{}l@{}}3. nancy pelosi and democrats block \textbf{bill} to \textbf{stop infanticide} for 75th time, refuse care for babies born alive\end{tabular} \\[0.1cm]
 &
  4. democrats block \textbf{bill} to \textbf{stop infanticide}. \\[0.1cm]
 &
  5. house democrats block request to vote on \textbf{bill} to \textbf{stop infanticide} \\ \bottomrule
\end{tabular}}
\label{tab:misinfo_narr_examples}
\end{table*}

\begin{table*}[!t]
\centering
\caption{\textbf{Potentially misleading narrative libraries.}}
\resizebox{0.9\textwidth}{!}{%
\begin{tabular}{@{}lllrr@{}}
\toprule
\multicolumn{1}{c}{\textbf{Source of texts}} &
  \multicolumn{1}{c}{\textbf{\begin{tabular}[c]{@{}c@{}}Potentially misleading \\ narratives library\end{tabular}}} &
  \multicolumn{1}{c}{\textbf{Description}} &
  \multicolumn{1}{c}{\textbf{\begin{tabular}[c]{@{}c@{}}Total number \\of narratives\\ (low-dimensional)\end{tabular}}} &
  \multicolumn{1}{l}{\textbf{\begin{tabular}[c]{@{}c@{}}Total number \\of narratives\\ (high-dimensional)\end{tabular}}} \\ \midrule
\multirow{3}{*}{\begin{tabular}[c]{@{}l@{}}Fake news \\ headlines \& \\ stories\\ ($30,120$ texts)\end{tabular}} &
  \begin{tabular}[c]{@{}l@{}}All fake-news \\ narratives\end{tabular} &
  \begin{tabular}[c]{@{}c@{}}All narratives found in fake news texts \end{tabular} &
  $209,271$ &
  $246,526$ \\
 &
  \begin{tabular}[c]{@{}l@{}}Recurring fake-news \\ narratives\end{tabular} &
  \begin{tabular}[c]{@{}c@{}}Subset of above, top $1\%$ in terms \\of frequency of occurrence \end{tabular} &
  $2,544$ &
  $2,600$ \\ \\
\begin{tabular}[c]{@{}l@{}}False fact-checked \\ claims\\ ($21,109$ texts)\end{tabular} &
  \begin{tabular}[c]{@{}l@{}}False-claims \\ narratives\end{tabular} &
  \begin{tabular}[c]{@{}c@{}}Narratives found in claims that  \\were fact-checked as false \end{tabular} &
  $10,838$ &
  $11,384$ \\ \bottomrule
\end{tabular}}
\label{tab:misinfo_narr_libs}
\end{table*}

\begin{table*}[!t]
\centering
\caption{\textbf{Results of our hypothesis tests.} These results show that across various categories of reliable mainstream news outlets and sets of potentially misleading narratives, potentially misleading narratives are present at a higher rate in co-shared articles than in the control group. $^{*}p<0.05$; $^{**}p<0.005$; $^{***}p<0.0005$  }
\resizebox{0.99\linewidth}{!}{%
\begin{tabular}{@{}lllrr@{}}
\toprule
\multicolumn{1}{c}{\textbf{\begin{tabular}[c]{@{}c@{}}Type of\\outlets\end{tabular}}} &
  \multicolumn{1}{c}{\textbf{\begin{tabular}[c]{@{}c@{}}Misinformation\\narrative\\library\end{tabular}}} &
  \multicolumn{1}{c}{\textbf{\begin{tabular}[c]{@{}c@{}}Dim. of \\ extracted \\narratives\end{tabular}}} &
  \multicolumn{1}{c}{\textbf{$n$}} &
  \multicolumn{1}{c}{\textbf{One-sided Wilcoxon-signed rank test results}} \\ \midrule[1pt]
\multirow{6}{*}{\begin{tabular}[c]{@{}l@{}}All reliable \\ outlets\end{tabular}} &
  \multirow{2}{*}{\begin{tabular}[c]{@{}l@{}}All fake news \\ narratives\end{tabular}} &
  Low-dim &
  $92577$ &
  statistic = 3029034904.0, $p = 0.0^{***}$, effect size statistic = $\infty$, 95\% confidence interval = ($4.9e^{-07}$, $4.9e^{-07}$) \\
 &
   &
  High-dim &
  $19845$ &
  statistic = 176302324.0, $p = 0.0^{***}$, effect size statistic = $\infty$, 95\% confidence interval = ($4.9e^{-07}$, $4.9e^{-07}$) \\[0.1cm]
 &
  \multirow{2}{*}{\begin{tabular}[c]{@{}l@{}}Recurring fake news \\ narratives\end{tabular}} &
  Low-dim &
  $1982$ &
  statistic = 1700135.0, $p = 8.73e^{-175^{***}}$, effect size statistic = 0.633, 95\% confidence interval = ($2.94e^{-06}$, $3.56e^{-06}$) \\
 &
   &
  High-dim &
  $657$ &
  statistic = 201765.0, $p = 3.3e^{-83^{***}}$, effect size statistic = 0.754, 95\% confidence interval = ($1.11e^{-06}$, $1.6e^{-06}$) \\[0.1cm]
 &
  \multirow{2}{*}{\begin{tabular}[c]{@{}l@{}}False-claims \\ narratives\end{tabular}} &
  Low-dim &
  $4413$ &
  statistic = 5132255.0, $p = 0.00095^{**}$, effect size statistic = 0.05, 95\% confidence interval = ($4.9e^{-07}$, $4.9e^{-07}$) \\
 &
   &
  High-dim &
  $318$ &
  statistic = 32746.0, $p = 2.97e^{-06^{***}}$, effect size statistic = 0.262, 95\% confidence interval = ($4.9e^{-07}$, $4.9e^{-07}$) \\[0.1cm] \hline
\multirow{6}{*}{\begin{tabular}[c]{@{}l@{}}Trustworthy\\outlets\end{tabular}} &
  \multirow{2}{*}{\begin{tabular}[c]{@{}l@{}}All fake news \\ narratives\end{tabular}} &
  Low-dim &
  $86871$ &
  statistic = 2533598901.0, $p = 0.0^{***}$, effect size statistic = $\infty$, 95\% confidence interval = ($5.62e^{-07}$, $5.62e^{-07}$) \\
 &
   &
  High-dim &
  $15912$ &
  statistic = 109142619.0, $p = 0.0^{***}$, effect size statistic = $\infty$, 95\% confidence interval = ($5.62e^{-07}$, $5.62e^{-07}$) \\[0.1cm]
 &
  \multirow{2}{*}{\begin{tabular}[c]{@{}l@{}}Recurring fake news \\ narratives\end{tabular}} &
  Low-dim &
  $1924$ &
  statistic = 1587705.0, $p = 1.09e^{-162^{***}}$, effect size statistic = 0.62, 95\% confidence interval = ($2.76e^{-06}$, $3.37e^{-06}$) \\
 &
   &
  High-dim &
  $587$ &
  statistic = 159336.0, $p = 3.09e^{-71^{***}}$, effect size statistic = 0.737, 95\% confidence interval = ($1.12e^{-06}$, $1.69e^{-06}$) \\[0.1cm]
 &
  \multirow{2}{*}{\begin{tabular}[c]{@{}l@{}}False-claims\\ narratives\end{tabular}} &
  Low-dim &
  $4281$ &
  statistic = 4482215.0, $p = 0.89$, effect size statistic = 0.002, 95\% confidence interval = ($4.68e^{-07}$, $5.62e^{-07}$) \\
 &
   &
  High-dim &
  $306$ &
  statistic = 29136.0, $p = 0.00012^{***}$, effect size statistic = 0.22, 95\% confidence interval = ($5.62e^{-07}$, $5.62e^{-07}$) \\[0.1cm] \hline
\multirow{6}{*}{\begin{tabular}[c]{@{}l@{}}Liberal\\outlets\end{tabular}} &
  \multirow{2}{*}{\begin{tabular}[c]{@{}l@{}}All fake news \\ narratives\end{tabular}} &
  Low-dim &
  $71654$ &
  statistic = 1831440473.0, $p = 0.0^{***}$, effect size statistic = $\infty$, 95\% confidence interval = ($9.2e^{-07}$, $9.2e^{-07}$) \\
 &
   &
  High-dim &
  $8235$ &
  statistic = 29733384.0, $p = 0.0^{***}$, effect size statistic = $\infty$, 95\% confidence interval = ($9.2e^{-07}$, $9.2e^{-07}$) \\[0.1cm]
 &
  \multirow{2}{*}{\begin{tabular}[c]{@{}l@{}}Recurring fake news \\ narratives\end{tabular}} &
  Low-dim &
  $1856$ &
  statistic = 1439940.0, $p = 9.26e^{-139^{***}}$, effect size statistic = 0.582, 95\% confidence interval = ($2.76e^{-06}$, $2.83e^{-06}$) \\
 &
   &
  High-dim &
  $396$ &
  statistic = 70018.0, $p = 7.59e^{-42^{***}}$, effect size statistic = 0.681, 95\% confidence interval = ($1.84e^{-06}$, $1.85e^{-06}$) \\[0.1cm]
 &
  \multirow{2}{*}{\begin{tabular}[c]{@{}l@{}}False-claims \\ narratives\end{tabular}} &
  Low-dim &
  $3956$ &
  statistic = 4442138.0, $p = 8.41e^{-14^{***}}$, effect size statistic = 0.119, 95\% confidence interval = ($-8.998e^{-07}$, $-8.65e^{-07}$) \\
 &
   &
  High-dim &
  $263$ &
  statistic = 24814.0, $p = 6.37e^{-10^{***}}$, effect size statistic = 0.381, 95\% confidence interval = ($9.2e^{-07}$, $9.2e^{-07}$) \\ \bottomrule
\end{tabular}}
\label{tab:misinfo_results_sig_testing}
\end{table*}

\begin{table*}[!t]
\centering
\caption{\textbf{Quantifying presence of potentially misleading narratives in the two groups of mainstream articles.} On average, $2.2\%$ of all narratives in a co-shared article are present in all narratives extracted across all misinformation content (recurring across fake news or present in fact-checked false claims), whereas $1.3\%$ of all narratives in a control article are present in all narratives extracted across all misinformation content (recurring across fake news or present in fact-checked false claims).}
\resizebox{0.8\linewidth}{!}{%
\begin{tabular}{@{}lr@{}}
\toprule
\multicolumn{1}{c}{} &
  \multicolumn{1}{c}{\textbf{\begin{tabular}[c]{@{}c@{}}\% of narratives in an article that are present \\ in the potentially misleading narrative set \\(average (standard deviation))\end{tabular}}} \\ \midrule
Co-shared mainstream news group &
  $2.2\% (4.2)$ \\
Control mainstream news group &
  $1.3\% (3.3)$ \\ \bottomrule
\end{tabular}}
\label{tab:misinfo_quant_presence_coshared_control}
\end{table*}

\clearpage

\section*{Figures}

\clearpage

\nocite{marquaridt1970generalized}
\nocite{akinwande2015variance}
\nocite{akaike1974new}
\nocite{burnham2004multimodel}
\nocite{AICcmodavg-package}
\nocite{bird2009steven}
\nocite{mann1947u}


\end{document}